\documentclass[11pt,english]{article}
\usepackage{rotating}
\usepackage{latexsym}
\usepackage{graphicx}
\usepackage{psfrag}
\usepackage{amsmath,amssymb}
\makeatletter
\def\section{\@startsection {section}{1}{\z@}{-3.5ex plus -1ex minus
 -.2ex}{2.3ex plus .2ex}{\large\bf}}
\def\subsection{\@startsection{subsection}{2}{\z@}{-3.25ex plus -1ex
minus -.2ex}{1.5ex plus .2ex}{\normalsize\bf}}
\makeatother
\makeatletter

\@addtoreset{equation}{section}

\makeatother

\newcommand{\captionfonts}{\small}
\makeatletter  
\long\def\@makecaption#1#2{%
  \vskip\abovecaptionskip
  \sbox\@tempboxa{{\captionfonts #1: #2}}%
  \ifdim \wd\@tempboxa >\hsize
    {\captionfonts #1: #2\par}
  \else
    \hbox to\hsize{\hfil\box\@tempboxa\hfil}%
  \fi
  \vskip\belowcaptionskip}
\makeatother   
\def\dslash{\raisebox{1pt}{$\slash$} \hspace{-6pt} \partial}
\def\pslash{\raisebox{1pt}{$\slash$} \hspace{-7pt} p}

\def\Aslash{\hspace{3pt}\raisebox{1pt}{$\slash$} \hspace{-9pt} A}
\def\Cslash{\hspace{3pt}\raisebox{1pt}{$\slash$} \hspace{-8pt} C}
\def\Dslash{\hspace{3pt}\raisebox{1pt}{$\slash$} \hspace{-8pt} D}

\def\bea{\begin{eqnarray}} \def\eea{\end{eqnarray}}
\def\be{\begin{equation}} \def\ee{\end{equation}} \def\nn{\nonumber}
\def\a{& \hspace{-7pt}}  \def\Z{{\bf Z}}
  
\def\ds{\displaystyle}

\def\ds{\displaystyle} 

\newcommand{\promille}{%
  \relax\ifmmode\promillezeichen
        \else\leavevmode\(\mathsurround=0pt\promillezeichen\)\fi}
\newcommand{\promillezeichen}{%
  \kern-.05em%
  \raise.5ex\hbox{\the\scriptfont0 0}%
  \kern-.15em/\kern-.15em%
  \lower.25ex\hbox{\the\scriptfont0 00}}

\begin{document}

\thispagestyle{empty}

\begin{center}
\hfill SISSA-61/2009/EP \\

\begin{center}

\vspace{1.7cm}

{\LARGE\bf Holographic Methods and Gauge-Higgs \\[3mm]
Unification in Flat Extra Dimensions}

\end{center}

\vspace{1.4cm}

{\bf Marco Serone}\\

\vspace{1.2cm}

{\em SISSA and INFN, Via Beirut 2-4, I-34151 Trieste, Italy}

\end{center}

\vspace{0.8cm}

\centerline{\bf Abstract}
\vspace{2 mm}
\begin{quote}\small

I review the holographic techniques used to efficiently study models with Gauge-Higgs Unification (GHU) in one extra dimension.  
The general features of GHU models in flat extra dimensions are then reviewed, emphasizing the aspects related to electroweak symmetry breaking. Two potentially realistic models, based on $SU(3)$ and $SO(5)$ electroweak gauge groups, respectively,  are constructed. 

\end{quote}

\vfill

\newpage

\tableofcontents

\section{Introduction}

Quantum field theories in more than four space-time dimensions have received a lot of attention in the past ten years. These theories are necessarily non-renormalizable and requires an ultra-violet (UV) completion, but they can admit an energy range where they are 
trustable low-energy effective theories with small UV-dependence. Model building in an effective bottom-up approach makes then sense in such theories. 

Extra dimensions allow us to address standard well-known problems in four-dimensional (4D) physics, such as the gauge  hierarchy problem, just to mention a very relevant example in particle physics, from a different perspective. The simple natural assumption of locality in the extra dimensions leads to striking solutions of the above problem, such as the possibility of having a fundamental TeV-sized quantum 
gravity scale \cite{ArkaniHamed:1998rs} or a TeV scale naturally generated by an extreme red-shift effect from a warped extra dimension \cite{Randall:1999ee}.  Higher-dimensional theories also open up the possibility of identifying the Standard Model (SM) Higgs boson as a gauge field polarization in the internal dimensions \cite{early,hos}.
With only one extra dimension, gauge invariance and locality imply that no divergencies can occur
in the Higgs effective potential, so that the gauge hierarchy problem is technically solved.
This idea, nowadays commonly denoted by Gauge-Higgs Unification (GHU),
including all its subsequent incarnations, plays or has played a crucial aspect  in some of the most promising models of new physics beyond the SM, alternative to supersymmetry. The deconstruction of 6D GHU models in flat space led to the development of little Higgs models \cite{ArkaniHamed:2001nc}, while GHU models in 5D warped space \cite{Agashe:2004rs} led to holographic duals of realistic composite Higgs models \cite{Kaplan:1983fs}.  Realistic five-dimensional GHU models
can also be constructed in flat space \cite{Panico:2006em}, although minimal models turn out to predict too light top and Higgs field and a too low new physics scale \cite{Scrucca:2003ra}. 
Interestingly enough, realistic GHU models, in both flat and warped space, naturally implement the idea of \cite{ArkaniHamed:1999dc} of effective Yukawa couplings suppressed by the geometry of the internal space.  Obvious constraints coming from the universality of the SM gauge interactions are satisfied by localizing all the SM fermions  (with the exception of the top and, to some extent, of the bottom quark) in the same region in the internal space. It is then convenient to use as fundamental low energy 4D fields the value of the bulk 5D fields at the (approximate) point in the internal space where the light fermions sit. In this field basis, the SM universailty of the couplings is manifest by construction and most of the new physics effects are encoded in universal parameters such as $S$ and $T$ \cite{Peskin:1991sw,Barbieri:2004qk}. This ``holographic basis'' \cite{Luty:2003vm,Barbieri:2003pr}
turns out to be particularly useful in GHU models, since it allows for a very efficient way to compute the Higgs potential, which is radiatively generated and calculable.

Aim of this work is  to give a rather pedagogical review to the main underlying features of (non-supersymmetric) GHU models in flat extra dimensions, using the holographic method mentioned before. 
I will mostly consider theories with just one extra dimension, because realistic  models have been constructed in this case only.  
Several considerations, based on symmetry arguments only, will not depend on the curvature of the extra dimension.
In fact, there is really not a fundamental difference between models defined in warped and flat extra dimensions, if one is interested to the LHC physics at the TeV scale.
A sub-class of 5D models in flat space with large localized gauge kinetic terms, indeed, seem to mimic 
all the main features of warped space models, with the additional advantage of being technically much easier to handle. We will see an example of this sort by constructing a 5D version in flat space of a certain warped space model \cite{Contino:2006qr}.

Flavour and CP issues will not be considered in this review.
An effective theory valid slightly above the TeV scale cannot actually address flavour problems, that involve much higher scales.\footnote{Warped models are expected to have even a lower cut-off than theories in flat space
 (see e.g. \cite{Haba:2009hw} for a recent comparison) but,  due to the warping, the cut-off scale depends on the position in the internal dimension. By locality, then, the effective cut-off for light fields can be  way much heavier than TeV (up to the Planck scale), so that flavour issues can be addressed.} On the other hand, assuming uncalculable UV corrections are under control (say, by a partial UV completion given by an underlying warped space model for the light generations), preliminary rough estimates on calculable corrections  show that no fundamental flavour problem seems to arise in models with flat extra dimensions \cite{Panico:2005dh}.  Similarly, the potential collider signatures of these models will not be addressed.

The review is organized as follows. In section 2 the holographic method to study theories with one extra dimension with boundaries (i.e. an interval) is introduced; in subsection 2.1 it is extended to fermions and in subsection 2.2 to gauge fields. In subsection 2.3 the universal parameters of \cite{Barbieri:2004qk} are introduced and a potential problem affecting the $Z\bar b_L b_L$ coupling in GHU models presented.  In section 3 the main features of GHU models are presented, with an explicit derivation of the Higgs potential and Yukawa coupling in a simple toy model.
In section 4 two realistic models are presented. In subsection 4.1 a model based on an $SU(3)$ electroweak gauge group where Lorentz symmetry is broken along the internal dimension \cite{Panico:2005dh} is rewieved; in subsection 4.2 a model based on an $SO(5)$ electroweak group with large localized gauge kinetic terms, mimicking its warped space relative \cite{Contino:2006qr}, is constructed and briefly analyzed; in subsection 4.3 we give a brief overview of GHU models in more than one extra dimension. I conclude in section 5. I summarize in Appendix A the conventions used in the text and report some technical details in Appendices B and C.

Although I have tried to be as pedagogical as possible, due to lack of space  
I have not included in this review basic general aspects about theories in extra dimensions, which are then assumed to be vaguely familiar to the reader. I refer the interested reader to the excellent reviews \cite{Rubakov:2001kp,Sundrum:2005jf,Rattazzi:2003ea} for a an overview on the general theoretical aspects of theories in extra dimensions, seen from a wider perspective.

\section{Holographic Description of 5D Field Theories on an Interval}

The standard procedure to derive a low-energy effective Lagrangian describing the 
massless excitations of higher dimensional fields is the Kaluza-Klein (KK) reduction,
in which one writes the higher dimensional fields as
\be
\Phi(x,y)=\sum_n \phi_n(x) f_n(y)\,,
\label{PhiKK}
\ee 
where $f_n(y)$ are eigenfunctions in the internal space and $\phi_n(x)$ the corresponding 4D fields, 
associated with canonical states with mass $M_n$.
In bottom-up approaches to 5D model building when the extra dimension is an interval $I$ (or equivalently the orbifold
$S^1/\Z_2$),  finding the spectrum of the KK resonances is not always straightforward,
even in flat space, due to the fact that the most generic action one can write is of the form
\be
S = \int \! d^4x \int_0^L\! dy {\cal \, L}
\ee
with ${\cal L}$ the following Lagrangian:
\be
{\cal L} = {\cal L}_5(y) +2 \delta(y) {\cal L}_0  +2 \delta(y-L) {\cal L}_L\,.
\ee
The Lagrangian terms  ${\cal L}_5$, ${\cal L}_0$ and ${\cal L}_L$ contain the most general set of operators 
compatible with the (global and local) symmetries of the theory up to a given dimensionality.
At the action level 5D Poincar\`e symmetry is always broken by the form of the space-time geometry $R^4\times I$.
Far away from the fixed-points, however, the theory locally looks like $R^5$ and hence ${\cal L}_5$ should be 5D Poincar\`e invariant. 
On the the hand, ${\cal L}_{0,L}$ manifestly break the 5D Poincar\`e symmetries to its 4D subgroup.
In presence of ${\cal L}_{0,L}$, finding the spectrum of the KK states, although conceptually easy, can be technically quite hard. 
In these situations an alternative approach can be used, where all the information of the theory is encoded in the values of the 5D fields at just one end-point of the segment. For this reason, this approach can be called ``holographic''. Choosing, say, $y=0$ as end-point, we define the holographic field 
\be
\hat \Phi(x) \equiv \Phi(x,y=0)\,.
\label{BoundaryDef}
\ee
For all states $n$ such that 
$\langle 0| \hat\Phi(x) |n \rangle \neq 0$ or, stated in other words, $f_n(0)\neq 0$, $\hat\Phi$ is a good
interpolating field. The simplest possible example one can consider is given by
a free scalar field with $(++)$ boundary conditions (b.c.) and ${\cal L}_{0,L}=0$: 
\be
{\cal L}_5 = \frac 12 (\partial_M \Phi)(\partial^M \Phi) \,, \ \ \ \ \partial_y \Phi(x,0) =\partial_y \Phi(x,L) = 0\,.
\label{LfreePhi}
\ee
Using eq.(\ref{PhiKK}), the 5D Klein-Gordon equation admits solutions of the form
\be
f_n(y) = A_n e^{i M_n y} + B_n e^{-i M_n y}\,,
\ee
where $A_n$ and $B_n$ are integration constants and $M_n$ are the mass eigenstates.
The b.c.  fix the $f_n$'s and the masses $M_n$ to be
\be
f_n(y)  =  \frac{2^{(1-\delta_{n,0})/2}}{\sqrt{L}}\cos\Big(\frac{\pi n y}{L}\Big) \,, \ \ \   M_n = \frac{\pi n}{L}\,,  \ \  n =0,1,\ldots,\infty\,.
\ee
The 4D Lagrangian is 
\be
{\cal L}^{KK} = \int_0^L\! dy \,{\cal L}_5 = \frac 12  \sum_{n=0}^\infty\Big[ (\partial_\mu \phi_n)^2 - M_n^2 \phi_n^2 \Big]\,.
\ee
Let us now turn to the holographic approach. It is convenient here to adopt  a mixed basis, with a description in terms of momenta in 4D
and of physical space in the internal dimension, so that we write $\Box_5 = -p^2 - \partial_y^2$, with $p^2 = p_\mu p^\mu =p_0^2- \vec{p}^{\,2}$. 

In the holographic approach, the b.c. at $y=0$ is replaced by the definition of the boundary field (\ref{BoundaryDef}). 
Given $\hat \Phi(x)$ and the b.c. at $y=L$, the bulk equations of motion (e.o.m.) admit a unique solution. No boundary term at $y=0$ arises in varying the action $S=\int d^4x dy {\cal L}_5$, since the boundary field is taken fixed: $\delta \hat \Phi = 0$.
The most general solution to the 5D Klein-Gordon equation for $\Phi$ reads
\be
\Phi(p,y) = A(p) \cos(py) + B(p) \sin (p y)\,.
\ee
The b.c.  at $y=L$ fix  $B(p) = \tan (p L) A(p)$, with $A(p) = \hat \Phi(p)$,
using for simplicity the same letter for the Fourier transform $\Phi(p)$ of the field $\Phi(x)$.
The solution can be written as
\be
\Phi(p,y) = G_{++}(p,y)  \hat \Phi(p)\,,
\label{HolPhi}
\ee
with 
\be
G_{++}(p,y) =  \cos (p y) +\tan (p L) \sin(p y)\,.
\label{b2BProp}
\ee
Given a value of $\Phi(p)$ at the boundary $y=0$, there is a unique field extension in the bulk,
given by eq.(\ref{b2BProp}). The function $G_{++}(p , y)$ is called bulk-to-boundary propagator
for obvious reasons. The holographic 4D momentum Lagrangian ${\cal L}^H$ is obtained by plugging the solution (\ref{HolPhi})
back in eq.(\ref{LfreePhi}). It reads
\bea
{\cal L}_{++}^{H} &  = & \frac 12  \int_0^L dy \Big[ p^2 \Phi^2 - (\partial_y \Phi)^2 \Big]=\frac 12 \int_0^L dy \Big[ \Phi(p^2 +\partial_y^2) \Phi \Big] -\frac 12  \Big[ \Phi \partial_y \Phi \Big]^L_0 \nn \\
& = &\frac 12   \Phi \partial_y  \Phi(y=0) = \frac 12 
 p \tan (p L) \hat \Phi^2\,,
\label{L4holo}
\eea
where for simplicity of notation $\hat \Phi^2$ stands for $\hat \Phi(-p) \hat \Phi(p)$. The Lagrangian (\ref{L4holo})  contains an infinite sum of higher-derivative quadratic terms.
Despite the apparently non-local nature of $p = \sqrt{p_{\mu} p^\mu} = \sqrt{-\Box_4}$, all the terms arising from the expansion of the tangent are local. The single holographic field $\hat \Phi$ encodes all the KK states.
Their mass spectrum  is encoded in the zeros of a single function, $p\tan (pL)$. There is of course a linear relation between $\hat \Phi$ and the KK fields $\phi_n$:  $\hat \Phi =  \sum_{n=0}^\infty f_n(0) \phi_n$. Since the KK fields are orthonormal, the following relations between the momentum space propagators in the two approaches should hold:
\be
\frac{ \cot (pL)}{p}= 
 \frac 1L
\bigg[ \frac{1}{p^2} + 2 \sum_{n=1}^\infty \frac{1}{p^2-M_n^2} \bigg]\,, 
\label{1DimHolSum}
\ee
as can easily be checked. The key point of the holographic approach is that one can trade the orthonormal zero mode $\phi_0$ for $\hat \Phi$ as
effective low-energy field.

In a similar fashion, one might also define an holographic Lagrangian for other choices of b.c., like Neumann-Dirichlet $(+-)$, Dirichlet-Neumann $(-+)$ or Dirichlet-Dirichlet  $(- -)$. 
Contrary to the $(++)$ case, no massless mode arises for these choices of b.c and such Lagrangians should not be considered now as proper low-energy effective Lagrangians since the holographic field $\hat \Phi$ creates and destroys only massive KK particles. Yet, they make sense and take into account the effect of the KK states. For Dirichlet $(-)$ b.c. at $y=L$, the bulk-to-boundary propagator is
\be
G_{+-}(p,y) =  \cos (p y) -\cot (p L) \sin(py) \,.
\ee
The $(-+)$ and $(- -)$ b.c. dot not allow to choose the interpolating field as $\Phi(y=0)$,  since the latter identically vanishes.  This problem is easily solved by noticing that an effective $(-)$ b.c. can always be derived dynamically from a $(+)$ b.c. by introducing a localized large mass term $\Lambda$ at $y=0$.\footnote{Given the 5D mass dimensions of $\hat \Phi$ as implied by eq.(\ref{BoundaryDef}), this term
is actually of the form $\Lambda^2 L \hat \Phi^2/2$.}
When $\Lambda\rightarrow \infty$, the $(-)$ b.c. is recovered.
For completeness, we report below the holographic Lagrangians arising from all possible b.c., in presence also of a 5D bulk mass term $m$:
\bea
{\cal L}_{++}^{H} \a  = \a \frac 12  \omega \tan (\omega L) \hat \Phi^2, \ \ \ \ \ \ \ \ {\cal L}_{-+}^{H}   = \frac 12\Big(\omega \tan (\omega L)- \Lambda^2L \Big) \hat \Phi^2, 
\label{scalarHolo} \\
{\cal L}_{+-}^{H} \a  = \a -\frac 12   \omega \cot (\omega L) \hat \Phi^2,\  \  
\ \ \ \ 
{\cal L}_{--}^{H}   = - \frac 12\Big(\omega \cot (\omega L)+ \Lambda^2 L\Big) \hat \Phi^2, 
\eea
where $\omega = \sqrt{p^2-m^2}$.  In all cases, the zeros of the
inverse propagators  agree with the expected KK. masses when $\Lambda\rightarrow \infty$. Notice that the mass eigenvalues of a given b.c. at $y=0$ are essentially given by looking at the poles of the inverse propagator with the opposite b.c. at $y=0$.

The examples reported above are so simple that one does not actually gain much in using a holographic rather than a KK approach. However, it should be now clear that the situation changes if we add localized Lagrangian terms. In particular, the addition of ${\cal L}_0$ is quite harmless in the holographic approach, since it does not alter the 5D bulk equations of motion.
One simply sums it to the Lagrangian terms ${\cal L}^{H}$ found before. This is clear, considering that the holographic approach is  an effective method where one integrates fields values for $y\neq 0$ and this integration, by locality, is not altered by the addition of terms localized at $y=0$. On the contrary, in the KK approach one has to compute again the 5D wave functions $f_n$, perturbed
by the localized term ${\cal L}_0$. In a sufficiently complicated set-up, then, the computation of the mass spectrum is typically more easily performed in the holographic approach.
Trilinear or higher couplings can also be computed. The logic is the same. One solves the e.o.m. in the bulk, now in a series expansion in the couplings, and then plug the results back in the action.
Contrary to the quadratic case, the bulk terms no longer vanish and higher terms are obtained by 
explicitly performing the integral over the internal space. See \cite{Luty:2003vm} for more details and \cite{Panico:2006em,Agashe} for some explicit examples in flat and warped space, respectively.

It is important to stress that the holographic technique reviewed here, although the terminology used is often similar (holographic fields, bulk-to-boundary propagators) does not imply the existence of any supposed ``dual'' purely 4D theory, related by some sort of AdS/CFT correspondence \cite{Ads-cft,ArkaniHamed:2000ds}. It is just a technical device, as explained, to conveniently perform computations.
In warped space in a slice of $AdS_5$, like in the Randall-Sundrum models and generalizations thereof,
the situation is different, since one can, by a change of language, express all quantities computed in the 5D theory as quantities of a ``chiral Lagrangian"  supposed to be the low-energy theory of a (typically unknown)  dual 4D 
CFT with spontaneous breaking of the conformal symmetry in the IR.

The generalization of the holographic approach to more than one extra dimension is completely straightforward if the internal space is a direct product of a compact space (of dimension $d\geq 1$)
times an interval. The resulting theory would be described by a $3+d$ dimensional Lagrangian which is then studied by means of a standard KK procedure. If the internal space is compact and without boundaries, there is clearly not a sensible way to use the holographic approach, unless the space is singular (such as orbifolds), in which case one might define the holographic field at some orbifold singularity. In addition to possible subtleties related to the singularity itself, another general problem emerges, since the momentum space propagator 
will have classical divergencies, due to the multidimensional sum over the KK states appearing in the generalization of eq.(\ref{1DimHolSum}). The simple holographic approach introduced here is hence not useful in more than one extra dimension.

\subsection{Fermions}

The free manifestly hermitian Lagrangian for a bulk fermion is
\be
{\cal L}_\psi = \frac i2 \bar \psi \gamma^M \partial_M \psi - \frac i2 (\partial_M \bar \psi) \gamma^M \psi - m \bar \psi \psi\,.
\ee
Being the Dirac equation first order in derivatives, at each boundary only one b.c. for $\psi_L$ or $\psi_R$ 
is required, the other being fixed by consistency with the bulk e.o.m. (see e.g. \cite{Csaki:2005vy} for a detailed description of the allowed b.c. for a fermion on an interval and Appendix \ref{conventions} for our conventions). Here we follow \cite{Contino:2004vy} in showing how to construct an holographic Lagrangian for fermions. We define at $y=L$ as $(-)$ the Dirichlet boundary condition corresponding to a vanishing chiral fermion component, denoting by $(+)$ the b.c. fixed by the Dirac equation for the other chirality. Let us define the holographic field with, say,  the left-handed
component: $\psi_L(y=0) \equiv \chi_L$. Contrary to the bosonic case, even by taking $\delta \chi_L = 0$, the variation of the action does not vanish. Whereas at $y=L$ the $(-)$ b.c. for $\psi_L$ or $\psi_R$ are enough to make the variation vanishing, at $y=0$ we are left with 
\be
\delta \int d^4x dy {\cal L}_\psi = -\frac 12  \int d^4x \Big( \bar \psi_L \delta \psi_R + \delta \bar \psi_R \psi_L\Big) (y=0) \,.
\label{variation}
\ee
By keeping as holographic field $\psi_R(y=0)\equiv \chi_R$, we would get eq.(\ref{variation}) with $L\leftrightarrow R$, but with opposite sign. 
Requiring the action to be invariant under any variation obliges us to add a new term, localized at $y=0$, of the form
\bea
{\cal L}_{0}\a= \a\frac 12 \Big( \bar \psi_L \psi_R + \bar \psi_R \psi_L\Big) (y=0)\,, \ \ \ \ \ {\rm holographic \; field} \;\;  \chi_L\,,  \nn \\
{\cal L}_{0}\a= \a -\frac 12 \Big( \bar \psi_L \psi_R + \bar \psi_R \psi_L\Big) (y=0)\,, \ \ \  {\rm holographic \; field} \;\;  \chi_R \,.
\label{L0fer}
\eea
The Dirac equation for the two chiral fermion components reads, in a ($p,y$) mixed basis
\bea
\pslash \, \psi_R \a = \a ( \partial_y +m)\psi_L \,, \nn \\
\pslash \, \psi_L \a = \a ( -\partial_y +m)\psi_R \,, 
\label{Deq}
\eea
where $\pslash =  \gamma^\mu p_\mu$. It is straightforward to write the general solutions to the
Dirac equation (\ref{Deq}) in terms of $\chi_L$. Omitting, for simplicity of notation, the momentum dependence of all
quantities, one has
\be
\left\{
\begin{array}{l}
\ds\psi_L(y)= \frac{G_+(y,m)}{G_+(m)} \chi_L \,, \\
\ds\psi_R(y)=\frac{G_-(y,m)}{G_+(m)}\frac{\pslash}p \chi_L\,,
\end{array}
\right.\!\! \psi_R(L)=0, \ \  
\left\{
\begin{array}{l}
\ds\psi_L(y)= \frac{G_-(y,m)}{G_-(m)} \chi_L \,, \\
\ds\psi_R(y)=-\frac{G_+(y,-m)}{G_-(m)} \frac{\pslash}p\chi_L\,,
\end{array}
\right. \!\!\psi_L(L)=0,
\label{GfermionchiLD0}
\ee
with
\bea
G_{+}(y,m) \a = \a \omega \cos \omega (L-y) + m \sin \omega (L-y)\,, \ \ \ G_{+}(m)  \equiv  G_{+}(y=0,m)\ \nn \\
G_{-} (y,m) \a = \a p \sin \omega (L-y)\,,  \hspace{3.4cm}   G_{-}(m) \equiv G_{-}(m,y=0)\,.  
\eea
The solutions of the Dirac equations when we keep as holographic field $\chi_R$ are trivially deduced
from eq.(\ref{GfermionchiLD0}) by noticing that eqs.(\ref{Deq}) are invariant for
$L\rightarrow R$, $m\rightarrow -m$, $\pslash\rightarrow -\pslash$. Explicitly, we have
\be
\left\{
\begin{array}{l}
\ds\psi_R(y)= \frac{G_+(y,-m)}{G_+(-m)} \chi_R\,,  \\
\ds\psi_L(y)=-\frac{G_-(y,m)}{G_+(-m)}\frac{\pslash}p \chi_R\,,
\end{array}
\right.\!\! \psi_L(L)=0, \ \ 
\left\{
\begin{array}{l}
\ds\psi_R(y)= \frac{G_-(y,m)}{G_-(m)} \chi_R \,, \\
\ds\psi_L(y)=\frac{G_+(y,m)}{G_-(m)} \frac{\pslash}p\chi_R\,,
\end{array}
\right. \psi_R(L)=0\,,
\label{GfermionchiLD}
\ee
where we have used that $G_-(y,-m)=G_-(y,m)$.
The holographic Lagrangian, like in the scalar case, is given by plugging the classical solution back in the action.
The bulk action gives a vanishing contribution and only the localized term (\ref{L0fer}) matters. We get
\bea
{\cal L}_{L+}^H \a = \a \frac 12 \bar \psi \psi(0) =  \bar \chi_L \frac{G_-(m)}{G_+(m)} \frac{\pslash}p \chi_L\equiv \bar \chi_L \Pi^+_L(m) \frac{\pslash}p \chi_L
\,,\nn   \\
{\cal L}_{L-}^H \a = \a  \frac 12 \bar \psi \psi(0) = -\bar \chi_L \frac{G_+(-m)}{G_-(m)}  \frac{\pslash}p\chi_L\equiv \bar \chi_L \Pi^-_L(m) \frac{\pslash}p \chi_L
\,,  \nn  \\
{\cal L}_{R+}^H \a = \a  -\frac 12 \bar \psi \psi(0) = \bar \chi_R \frac{G_-(m)}{G_+(-m)} \frac{\pslash}p \chi_R\equiv \bar \chi_R \Pi^+_R(m)  \frac{\pslash}p\chi_R \,, \nn \\
{\cal L}_{R-}^H \a = \a - \frac 12 \bar \psi \psi(0) = -\bar \chi_R \frac{G_+(m)}{G_-(m)}  \frac{\pslash}p\chi_R\equiv \bar \chi_L \Pi^-_R(m)  \frac{\pslash}p\chi_R \,,
\label{LagHolFerm}
\eea
where $\pm$ stand for the b.c. at $y=L$ of the holograhic field which is retained. 
The obvious natural choice of chiral fermion component to be chosen as holographic field is  the one with $(+)$ b.c. at $y=0$, so that it is not trivially vanishing. Like in the scalar case, it is not difficult to see that the zeros $M_n$ of the inverse propagators appearing in eqs.(\ref{LagHolFerm})  coincide with the KK mass eigenvalues. Notice also the similarity between  eqs.(\ref{LagHolFerm}) and eqs.(\ref{scalarHolo}), in particular the identification of poles (zeros) of the inverse propagator with the zeros (poles) of the one with opposite holographic chirality component and b.c. at $y=L$.
The function $G_{-}(m)$ has also a zero at vanishing momentum, for any value of the bulk mass $m$, corresponding to a chiral zero mode. Depending on the sign of $m$,
the zero mode is exponentially localized at $y=0$ or $y=L$, as obvious from eq.(\ref{Deq})  when its left hand side vanishes. For $m=0$, the zero mode has a flat profile in the extra dimension. 
In certain circumstances, that we will extensively discuss later on,
it may be useful to keep the ``wrong'' $(-)$ chiral component. A way to implement its $(-)$ b.c. at $y=0$ is by introducing a fermion Lagrange multiplier
$\lambda$, with opposite chirality, and add to the Lagrangian the further localized term
\be
{\cal L}_{0,l.m.} = \bar \lambda \chi +\bar \chi \lambda\,.
\label{L0lm}
\ee 
Thanks to the term (\ref{L0lm}), the condition $\chi=0$ at $y=0$ dynamically arises from 
the e.o.m. of $\lambda$.  On the other hand, it we solve for $\chi$, we get an 
holographic action for $\lambda$, that becomes a good holographic field to describe the possible zero modes
coming from the $(+)$ chiral component that has been integrated out.

The holographic description can easily be extended to the
situation in which localized fermions mix with bulk fermions. 
For example, consider a left-handed chiral fermion $q_L$ localized at $y=0$, mixing
with the right-handed component of a bulk fermion $\psi_R$:
\be
{\cal L}_0 = \bar q_L i\dslash q_L + e (\bar q_L \psi_R + \bar \psi_R q_L)+\tilde{\cal L}_0\,,
\ee
where $e$ is the mixing parameter and $\tilde{\cal L}_0$ is a boundary term.
The b.c. for the bulk fermion at $y=0$ are clearly neither $(+)$ or $(-)$, due to the mixing. 
It is natural to choose the localized fermion $q_L$ as holographic field. 
Since $\delta \psi_L(0)\neq 0$, the boundary action is fixed by requiring the vanishing of the  
boundary variation of $\psi_L$, giving  $\tilde{\cal  L}_0 = -1/2 ( \bar \psi_L \psi_R + \bar \psi_R \psi_L) (0)$, as in eq.(\ref{L0fer}) but with opposite sign. 
The boundary variation of
$\psi_R$, instead, dynamically fixes
\be
\psi_L(0) = e q_L\,.
\ee
The bulk e.o.m. for $y\neq 0$, as well as the b.c. at $y=L$, are unaffected by the presence of the localized fermion $q_L$, so that all the bulk-to-boundary propagators
are the same as before. The holographic Lagrangian is easily found to be
\be
{\cal L}^H_{\pm} =  \bar q_L \,\pslash \Big(1+e^2 \frac{\Pi_L^\pm(m)}{p} \Big) q_L\,,  
\ee
where $\pm$ in ${\cal L}^H$ refers to the b.c. at $y=L$ of $\psi_L$.

It is natural to ask if there is a rationale to neglect all localized (mass and kinetic) terms for the bulk fermions (and in general for other fields) in model building in extra dimensions. The answer is yes, motivated by the fact that localized mass and kinetic terms are less relevant than the corresponding bulk mass and kinetic terms. Due to the lower dimensionality of the Lagrangian density, a localized mass term is effectively a coupling constant and, similarly, the coefficient multiplying a possible localized kinetic term would be irrelevant, of dimension $-1$. If one requires that localized mass or kinetic terms vanish at some scale, say the cut-off scale $\Lambda$, they will be radiatively generated \cite{Georgi:2000ks} but with a small coefficient.  In a low-energy effective field theory approach, it then makes sense to neglect them.\footnote{Among all possible localized operators, those with derivatives along the internal dimension require special care and are
more complicated to handle \cite{delAguila:2003bh}. It has been
pointed out in \cite{delAguila:2006kj} that their effect can however be eliminated
by suitable field redefinitions (see also \cite{Lewandowski:2001qp}).}
On the other hand, one can make use of such terms, if useful, by assuming that they do not vanish
at some scale, or even that they are large. We will  make use of localized fermion mass (and gauge kinetic) terms in a GHU model to be introduced later on. We refer the reader to Appendix B for an explicit derivation of  how localized mass terms at $y=L$ change the bulk-to-boundary fermion propagators.

\subsection{Gauge fields} 
\label{gaugeHOLO}

The holographic description for gauge fields follows along the same lines, but is slightly
complicated by the gauge-fixing procedure. 
Following \cite{Panico:2007qd}, we will consider a ``holographic gauge-fixing'' where  the 4D gauge fields are efficiently disentangled from the scalar degrees of freedom arising from their internal component in the extra dimension.
  
Consider a bulk Yang-Mills (YM) theory with group $G$ broken to $H$ at $y=L$ and to $H^\prime$ at $y=0$. We denote by $A_M^A$ the YM gauge field, where the superscript $A\in {\cal G}$ is the gauge index, which splits
into $A=(a,\hat a)$, with $a\in {\cal H}$ and $\hat a\in {\cal G}/{\cal H}$. The b.c. at $y=L$ for $A_M^A$ are the covariant
versions of the usual $(+)$ or $(-)$ b.c. for a scalar field:
\be
F_{\mu y}^a(y=L) = 0 \,, \ \ \ \ \ \ 
A_\mu^{\hat a}(y=L) = 0\,.
\label{Gaugebc}
\ee
Consistency with eq.(\ref{Gaugebc}) requires that the gauge parameters $\lambda^{\hat a}$ vanish at $y=L$, which is another way
of saying that at $y=L$ the group $G$ is broken to $H$.  Most of the degrees of freedom in the internal components of the gauge field $A_y^A$ can be gauged away. The $\xi$-gauges, canceling the mixing between $A_\mu$ and $A_y$ coming from the gauge kinetic term at quadratic level, are proportional to
\be
\frac{1}{\xi} (\partial_\mu A^\mu_A + \xi \partial_y A_y^A)^2\,.
\ee
The unitary gauge $\xi\rightarrow \infty$ gives $\partial_y A_y^A=0$, so that all the modes of $A_y^A$ can be gauged away, with the
exception of possible zero modes components, arising when $A_y^A$ has $(++)$ b.c., i.e. $A\in {\cal G}/{\cal H} \cap {\cal G}/{\cal H}^\prime$.  
When the number of physical scalar zero modes coming from $A_y$ are precisely ${\rm dim}\,(G/H)$ and no less (as will always be the case in the explicit models we will consider),  the simpler gauge $A_y^A=0$ can be taken by 
introducing extra degrees of freedom at $y=L$ and, at the same time, extra ${\rm dim}\,(G/H)$ 4D fields, so that the physical theory is left unchanged  \cite{Panico:2007qd}. The extra 4D fields $\pi_{\hat a}$ are encoded in the sigma-model field
\be
\Sigma(x) = \exp\Big[ i\frac{\pi_{\hat a}(x) t^{\hat a}}{f_\pi} \Big]\,.
\label{sigma}
\ee
We can use $\Sigma$ to make the b.c. (\ref{Gaugebc}) (which are only $H$-invariant) completely $G$-invariant, taking
\be
\big(F_{\mu y}^{(\Sigma^{-1})}\big)^a(y=L) = 0 \,, \ \ \ \ \ \ 
\big(A_\mu^{(\Sigma^{-1})}\big)^{\hat a}(y=L) = 0\,,
\label{GaugebcSigma}
\ee
where $A^{(g)}=g(A_M + i \partial_M ) g^\dagger$, $F^{(g)}=g F g^\dagger$ are the gauge transformed connection and field strength. 
Once restored the $G$ invariance of the b.c., the gauge choice $A_y=0$ can be taken. In this way, we have essentially traded the zero mode components of $A_y^{\hat a}$  for the fields $\pi_{\hat a}$. 
Under 4D gauge transformations at $y=L$, the fields $\pi_{\hat a}$ transform as $G/H$ Goldstone boson fields.

Let us now turn to the b.c. at $y=0$ and let us denote by $a^\prime \in {\cal H}^\prime$ and $\hat a^\prime\in {\cal G}/{\cal H}^\prime$ the unbroken and broken
generators there. We have $A_\mu^{\hat a^\prime}(y=0) = 0$ and $A_\mu^{a^\prime}(y=0)\equiv C_\mu^{a^\prime}$, the latter being 
identified as the holographic gauge fields.\footnote{If needed, instead of setting to zero $A_\mu^{\hat a^\prime}$ at $y=0$, in analogy to the scalar case discussed in 2.1, one can consider $(+)$ components for all the gauge fields at $y=0$ and dynamically get the $(-)$ b.c. by introducing mass terms of the form $\Lambda^2 (A_\mu^{\hat a^\prime})^2$, with $\Lambda\rightarrow \infty$.}  It is useful to perform a gauge transformation which brings back the b.c. at $y=L$
in the original form (\ref{Gaugebc}), giving now rotated b.c. at $y=0$:
\be
A_\mu(y=0)= C_\mu^{(\Sigma^{-1})}=\Sigma^\dagger (C_\mu+i\partial_\mu) \Sigma\,,
\label{BSigma}
\ee
where $C_\mu = C_\mu^{a^\prime} t^{a^\prime}$. 

The final result of this procedure is quite simple. In the gauge $A_y^A=0$, eqs.(\ref{Gaugebc}) turn to the standard $(-/+)$  b.c., $A_\mu^{\hat a}=\partial_y A_\mu^a=0$, and the Goldstone boson fields only appear at $y=0$ from eq.(\ref{BSigma}).
In this gauge the action reads, for each simple group factor, 
\be
S_{YM} = \frac{1}{g_5^2} \int \!d^4x dy\, {\rm Tr} \bigg[-\frac 12 F_{\mu\nu} F^{\mu\nu} + (\partial_y A_\mu)  (\partial_y A^\mu) \bigg]\,,
\ee
normalizing the generators as ${\rm Tr}\, t^A t^B = \delta^{AB}/2$ 
in the fundamental representation.
It is convenient to disentangle the transverse and longitudinal part of $A_\mu$. In momentum space
\be
A^\mu = \Big(\eta^{\mu\nu} - \frac{p^\mu p^\nu}{p^2} \Big) A_\nu +  \frac{p^\mu p^\nu}{p^2} A_\nu
 \equiv  (P^{\mu\nu}_t+P^{\mu\nu}_l) A_\nu = A^\mu_t + A^\mu_l \,.
 \ee
The e.o.m. for $A_\mu^t$ and $A_\mu^l$ read 
\be
(p^2+\partial_y^2) A_\mu^t = 0 \,, \ \ \ \ \ \ \partial_y^2 A_\mu^l = 0\,.
\ee
One easily finds
\bea
A_\mu^{a,t}(p,y)\a = \a G_g^{t,+}(p,y) A_\mu^{a,t}(p,0)\,, \ \ G_g^{t,+}(p,y) = \cos(py) + \tan(pL) \sin(py),\nn \\
A_\mu^{\hat a,t}(p,y) \a = \a G_g^{t,-}(p,y) A_\mu^{\hat a,t}(p,0) \,, \ \ G_g^{t,-}(p,y) = \cos(py) - \cot(pL) \sin(py), \nn \\
A_\mu^{a,l}(p,y)\a = \a G_g^{l,+}(p,y) A_\mu^{a,l}(p,0)\,, \ \ G_g^{l,+}(p,y) = 1,\nn \\
A_\mu^{\hat a,l}(p,y)\a = \a G_g^{l,-}(p,y) A_\mu^{\hat a,l}(p,0)\,, \ \ G_g^{l,-}(p,y) = 1-\frac yL.
\label{Ggauge}
\eea 
The holographic Lagrangian at quadratic level is given by 
\be
{\cal L}^H = -\frac{1}{g_5^2} {\rm Tr}\, \Big(A_\mu^t \partial_y A^{\mu,t}+A_\mu^l \partial_y A^{\mu,l} \Big)(y=0)\,,
\label{LholoG}
\ee
where eq.(\ref{BSigma}) has to be used to rewrite ${\cal L}^H$ in terms of $C_\mu$ and $\Sigma$.
The Lagrangian (\ref{LholoG}) is gauge invariant under $H^\prime$ local transformations, so that a residual 4D gauge-fixing has still to be imposed on $C_\mu$ to completely remove any gauge redundance. A useful choice is the Landau gauge $C_\mu^l =0$,
which removes mixing terms between $C_\mu$ and the Goldstone boson fields $\pi_{\hat a}$ at quadratic level.
The quadratic holographic Lagrangian in this gauge can easily be computed when $\langle  \pi_{\hat a} \rangle = 0$. One gets
\be
{\cal L}_{quad.}^H = \frac{1}{2g_5^2 Lf_\pi^2}p^2 \pi_{\hat a}^2 -\frac{P^{\mu\nu}_t }{2g_5^2} C_\mu^{a^{\prime}}\Pi_g^+(p) C_\nu^{a^{\prime}} -\frac{P^{\mu\nu}_t}{2g_5^2} C_\mu^{\hat a^{\prime\prime}} \Pi_g^-(p)  C_\nu^{\hat a^{\prime\prime}}\,,
\label{LholoGExp}
\ee
where 
\be
\Pi_g^+(p)=p \tan(pL), \ \ \ \  \Pi_g^-(p)= -p \cot(p L),
\label{PiGauge}
\ee
 $a^{\prime}\in {\cal H}^\prime \cap {\cal H}$, $\hat a^{\prime\prime}\in {\cal H}^\prime \cap {\cal G}/{\cal H}$. The kinetic term for the
Goldstone fields $\pi_{\hat a}$, the first term in eq.(\ref{LholoGExp}), arises from the longitudinal gauge field components 
and is the only non-vanishing contribution, at quadratic level, coming  from these components. 

\subsection{Universal parameters and $\delta g_b$}

In phenomenological models, the gauge fields $C_\mu^{a\prime}$  contain the SM gauge bosons.
More precisely, they can be identified with the SM gauge fields, provided that the SM fermions couple approximately in an universal way to them.
Along the lines of  \cite{Barbieri:2004qk}, which we closely follows here,
we can write the SM kinetic terms in the form (reabsorbing the gauge coupling contant in the form factors $\Pi$):
\be
 -P^{\mu\nu}_t \Big[ W_\mu^{+} \Pi_{W^+W^-}(p) W_\nu^- +\frac 12 W_\mu^{3} \Pi_{W_3W_3}(p) W_\nu^3  +\frac 12 B_\mu \Pi_{BB}(p) B_\nu +W_\mu^{3} \Pi_{W_3B}(p) B_\nu\Big],
\label{LagPi}
\ee
where $W$ and $B$ are the SM $SU(2)_L$ and $U(1)_Y$  gauge fields, respectively.  
By expanding in derivatives the four form factors appearing in (\ref{LagPi}), we get a series of higher dimensional operators, suppressed by the scale $1/L$.
Keeping terms up to quadratic order in $p^2$ would give 12 coefficients. 
Three of them define the 4D SM coupling constants $g\equiv g_4$, $g^\prime\equiv g_4^\prime$ and the Higgs vacuum expectation value (VEV):\footnote{Notice that our convention for the Higgs VEV differ by a $\sqrt{2}$ factor from that in \cite{Barbieri:2004qk}.}
\be
\frac{1}{g^2} = \Pi^\prime_{W^+W^-}(0)\,, \ \ \  \frac{1}{g^{\prime 2}} = \Pi^\prime_{BB}(0)\,, \ \ \ v^2 = -4\Pi_{W^+W^-} \approx (246\, {\rm GeV})^2\,,
\label{defggprime}
\ee
where a prime stands for a derivative with respect to $p^2$. In the above non-canonical basis, the conservation of the electromagnetic charge $Q=T_3+Y$   implies 
\be
\Pi_{W_3W_3}(0) +\Pi_{W_3 B}(0) = \Pi_{W_3W_3}(0) +2\Pi_{W_3 B}(0) +\Pi_{BB}(0)=0,
\ee
 so that only $12-3-2=7$ coefficients are independent.
In \cite{Barbieri:2004qk} they have been denoted by $\hat S$, $\hat T$, $\hat U$, $V$, $X$, $Y$ and $W$. The first three are rescaled versions of the 
Peskin-Takeuchi $S$, $T$ and $U$ parameters \cite{Peskin:1991sw}.  The higher dimensional operators that are more sensitive to new physics effects are \cite{Grinstein:1991cd}
\be
{\cal L}={\cal L}_{SM} + \frac{2}{v^2}\bigg[c_{WB}{\cal O}_{WB}+c_{H}{\cal O}_{H}+c_{WW}{\cal O}_{WW}+c_{BB}{\cal O}_{BB}\bigg]\,,
\ee
where
\bea
{\cal O}_{WB}\a = \a \frac{1}{g g^\prime} (H^\dagger \tau^a H) W_{\mu\nu}^a B^{\mu\nu}\,, \ \ \ 
{\cal O}_{H} = |H^\dagger D_\mu H|^2\,, \nn \\
{\cal O}_{WW} \a = \a \frac{1}{2g^2} (D_\rho W_{\mu\nu}^a)^2\,, \hspace{1.45cm}
{\cal O}_{BB} = \frac{1}{g^{\prime ^2}} (\partial_\rho B_{\mu\nu})^2\,.
\label{dim6op}
\eea
The parameters $\hat S$, $\hat T$, $W$ and $Y$ are defined and related to the coefficients of the operators (\ref{dim6op}) as follows:
\bea
\hat S \a\equiv\a g^2 \Pi^\prime_{W^3B}(0)=2\cot\theta_W c_{WB}\,, \hspace{.7cm}\ \hat T \equiv\frac{g^2 }{M_W^2}\Big[\Pi_{W_3W_3}(0)- \Pi_{W^+W^-}(0)\Big]=-c_H,\nn \\
W \a\equiv\a \frac 12 g^2 M_W^2  \Pi^{\prime\prime}_{W_3W_3}(0)=-g^2 c_{WW} \,, \ \ 
Y \equiv \frac 12 g^{\prime 2} M_W^2  \Pi^{\prime\prime}_{BB}(0)=-g^2 c_{BB}\,,
\label{STWYdef}
\eea
where $M_W=g v/2$ and $\theta_W$ is the SM weak-mixing angle.

The typical values of $\hat S$, $\hat T$, $W$ and $Y$ for two broad (unnatural) models of theories in extra dimensions can easily be computed.
Models where all fermions and the Higgs field are completely localized at $y=0$, while the SM gauge fields propagate in the bulk give
\be
\Pi_{W_a W_b}(p) =  \frac{\delta_{ab}}{g_5^2} p\tan(pL)-\frac{\delta_{ab}v^2}{4}\,,   \ \ \Pi_{BB}(p) = \frac{1}{(g_5^\prime)^2} p\tan(pL)-\frac{v^2}{4}\,,
\ \  \Pi_{W_3B} = \frac{v^2}{4}\,,
\label{STWY}
\ee
where the $v^2$ terms in (\ref{STWY}) are the trivial contributions due to the localized Higgs field. Eqs.(\ref{defggprime}) and (\ref{STWYdef}) 
quickly give
\be
\frac{L}{g_5^2} = \frac{1}{g^2}\,, \ \ \ \frac{L}{(g_5^\prime)^2} = \frac{1}{(g^\prime)^2}
\ee
and 
\be 
\hat S = \hat T = 0\,, \ \ \ \ W = Y = \frac 13 m_W^2 L^2\,.
\ee
When the Higgs is a bulk field, with fermions still localized at $y=0$, the bulk e.o.m. for the SM fields are changed
by the bulk Higgs contribution that reads (suppressing the Lorentz indices)
\be
{\cal L}_{Higgs} \supset  \frac{v^2}{8L} \Big[W_1^2 + W_2^2+(W_3-B)^2\Big]\,.
\label{HiggsBulk}
\ee
Correspondingly, we now have 
\be
\Pi_{W_1W_1} = \Pi_{W_2W_2} = \frac{1}{g_5^2} \omega \tan(\omega L)\,.
\label{Pi11}
\ee
with $\omega=\sqrt{p^2-g_5^2 v^2/(4L)}$. 
The computation of the remaining from factors is best done by going to the $Z$, $\gamma$ basis,
$Z=B-W_3$, $\gamma=(g_5/g_5^\prime) B + (g_5^\prime/g_5)  W_3$ and then back to $W_3$ and $B$.
One gets
\be
\Pi_{BB} =  \frac{g_5^2}{g_5^{\prime 2}} \Pi_{\gamma\gamma}+\Pi_{ZZ}\,, \ \ \ 
\Pi_{BW_3} = \Pi_{\gamma\gamma}-\Pi_{ZZ}\,, \ \ \ 
\Pi_{W_3W_3}= \frac{g_5^{\prime 2}}{g_5^2} \Pi_{\gamma\gamma}+\Pi_{ZZ}\,,
\label{gammaZBW}
\ee
with
\be
\Pi_{\gamma\gamma} =  \frac{ p \tan(pL)}{g_5^2+g_5^{\prime 2}}\,,  
  \ \ \ \ \Pi_{ZZ} = \frac{ \tilde\omega \tan(\tilde \omega L)}{g_5^2+g_5^{\prime 2}}\,, \ \ 
  \tilde\omega=\sqrt{p^2-\frac{(g_5^2+g_5^{\prime 2})v^2}{4L}}\,.
  \label{PigammaZ}
\ee
Using eqs.(\ref{Pi11})-(\ref{PigammaZ}) and neglecting ${\cal O}(M_W^4 L^4)$ corrections, one easily finds
\be
\hat S = \frac 23 M_W^2 L^2 \,, \ \ \ 
\hat T = \frac 13 \tan^2\theta_W M_W^2 L^2\,,\ \ \ 
W =  \frac 13 M_W^2 L^2\,, \ \ \ 
Y =  \frac 13 M_W^2 L^2\,.
\label{stwyHiggsbulk}
\ee
The remaining 3 parameters $\hat U$, $V$ and $X$ are vanishing at this order.
It is clear from these two simple examples that  the new parameters $W$ and $Y$ have to be taken into account and cannot in general
be neglected in 5D model building. The above universal parameters also receive radiative corrections
from usual SM corrections, which have to properly be considered in performing fit with the data.
One should also pay attention on the possibility, not always negligible (see e.g. \cite{Barbieri:2007bh}), that new physics may significantly alter some SM not well measured or yet unknown couplings  (such as top or Higgs couplings) 
which then changes the SM corrections in a non-negligible way.
 
We have been focusing so far on universal corrections, but  new physics would in general affect fermions
in a species dependent way. Even neglecting flavour changing and CP violation effects, which will not be treated here, it has been shown
in \cite{Cacciapaglia:2006pk} that, aside from the universal parameters considered before, 3 other operators are particularly
sensitive to effects of new physics. They are parametrized by the distortion $\delta
g_b\equiv g_b-(g_b)_{SM}$ (or the $\epsilon_b$ parameter \cite{Altarelli:1993sz}) of the $Z\, b_L \overline b_L$ coupling
and by other two parameters which describe the deviation of the up and down quark couplings to the $Z$ boson.
The holographic approach allows to efficiently compute such corrections.
In order to illustrate the idea, we can consider a simplified situation of a bulk fermion with mass $m$ coupled
to an unbroken $U(1)$ gauge field $A$. By gauge invariance, we now clearly have $\delta g=0$, yet we can
compute $\delta g$ as a function of the gauge field momentum, in other words as a form factor.
Let us take $\psi_R(L)=\psi_R(0)=0$, $(++)$ b.c. for $A$ and keep $\psi_L(0)=\chi_L$ as holographic field.
The relevant coupling is the cubic interaction term
\be
{\cal L}^{(3)} = g_5 \int^L_0\!dy\, \overline \psi (p+q,y)  \Aslash(q,y) \psi(p,y) \,.
\label{3-vertex}
\ee
As further simplification, let us consider the kinematic
configuration in which $p^2=(p+q)^2=0$, and 
$q^2\ll m^2$.
By using the fermion and gauge bulk-to-boundary propagators (\ref{GfermionchiLD0}) and (\ref{Ggauge}), one easily computes the integral
over the internal coordinate in (\ref{3-vertex}). Keeping up to ${\cal O}(q^2)$ terms,  and adding the quadratic terms, we have
\be
{\cal L}_H =  \bar \chi_L\, \pslash \, \chi_L  - \frac{1}{2} q^2 C_\mu^t C^{\mu,t} + g
\bigg[1 + \frac{q^2}{m^2} F(m L)\bigg] \overline \chi_L(p+q) \Cslash^t(q) 
\chi_R(p) \,,
\label{3-vertex-corr}
\ee
where we have defined the 4D coupling $g=g_5 \sqrt{L}$ and rescaled $\chi_L \rightarrow \chi_L/\sqrt{Z_\chi}$,
$C_\mu\rightarrow C_\mu/\sqrt{L}$ to get canonically normalized fields, with
$Z_\chi=[m(\coth(mL)+1)]^{-1}$. 
The function $F$ in (\ref{3-vertex-corr})  is defined as
\begin{equation}
F(x) \equiv \frac 14 \bigg[ (1-x)(x\coth x-1) +x^2\bigg]\,.
\label{Fx-Zbb}
\end{equation}
As expected, at $q^2=0$, $\delta g =0$ by gauge invariance. 
At quadratic order in the gauge boson momentum, however, we get 
\begin{equation}
\frac{\delta g}{g} =\frac{q^2}{m^2} F(mL)\,.
\label{deltagU(1)}
\end{equation}
The corrections of the form (\ref{deltagU(1)}) are essentially unavoidable
for partially delocalized fields, which couple to the ``massive'' gauge fields $A_\mu(y)$, with
$y\neq0$. The typical size of deviations in the SM $Z\psi \bar \psi$ coupling, for
SM fermions identified as zero modes like $\chi_L$ above, are given by eq.(\ref{deltagU(1)}) with $q^2 \sim M_Z^2$.
As we will later see, the Yukawa couplings of the fermions $\chi$
are of order $mL/\sinh(mL)$, implying that for light SM fermions one has $m\gtrsim {\cal O}(10)/L$. For such values of $m$ and $1/L\sim$  TeV, the
SM coupling deviations for light fields, as given by eq.(\ref{deltagU(1)}),  are $\delta g_l/g_l \sim 10^{-4}$, below current experimental bounds.
The situation is different for heavy fermions, in particular for the left-handed bottom quark $b_L$. Being related by $SU(2)_L$ to  the top quark $t_L$,
$b_L$ has to have a partial delocalization in the bulk which, in the illustrative model above, means $m\sim {\cal O}(1/L)$.
For such values of $m$, eq.(\ref{deltagU(1)}) gives $\delta g_b/g_b \sim 10^{-3}$, which is on the edge of current experimental bounds. In more complicated situations, in addition to the correction (\ref{deltagU(1)}), other corrections can appear, coming from the mixing, after electroweak symmetry breaking (EWSB),  
of fermions in different representations of $SU(2)_L$.  Luckily enough, these corrections, which might be quite large, can be significantly reduced by imposing certain discrete symmetries in models with a custodial $SU(2)$ symmetry  \cite{Agashe:2006at}.

Summarizing, in GHU models the most significant flavour and CP conserving bounds arise
from the universal parameters $\hat S$, $\hat T$, $W$, $Y$ and the coupling deviation $\delta g_b$.

\section{Gauge-Higgs Unification}

Gauge-Higgs Unification (GHU) is an acronym which encodes all models in extra dimensions
where the SM Higgs boson $H$ is identified with the zero mode of an internal component of a higher-dimensional gauge field.
By choosing suitable gauge groups in the extra dimensions, one then incorporates
all SM gauge bosons ($\gamma$, $W^{\pm}$, $Z$ and gluons) and the Higgs field $H$
as arising from different components of the same higher dimensional gauge field $A_M$.
Since $H$ is a doublet under $SU(2)_L$, GHU models necessarily require gauge groups
$G\supset G_{SM}=SU(3)_c\times SU(2)_L\times U(1)_Y$.
Most (semi)-realistic GHU models are defined in 5 or 6 dimensions.
In the following we will mainly consider the (more promising) GHU models in 5D, 
briefly reviewing 6D constructions later on.

In GHU models in 5D, the group $G$ is chosen such that under the decomposition $G\rightarrow G_{SM}$,
some Goldstone fields  $\pi_{\hat a}$ appearing in eq.(\ref{sigma}) have the correct quantum numbers to be identified with $H$.
The key idea of GHU models is that the Higgs field, being the component of a gauge field,
is protected by radiative quadratic divergencies by the underlying higher-dimensional
gauge symmetry. In fact, gauge invariance forbids any local potential for $H$ in the interior of the segment
(bulk), the only allowed gauge-invariant local operators being built with the field strength $F_{MN}$.
This is particularly clear in the holographic approach where, as we have just seen, there is a gauge in which 
$H$ does not appear at all in the bulk! 
The non-linear symmetry transformations
\be
\delta \pi_{\hat a} = \lambda_{\hat a} + \ldots 
\label{nonlinear}
\ee
forbid the appearance of any local potential for $\pi_{\hat a}$ at the boundaries as well.
The Higgs potential $V(H)$ in 5D GHU models is hence necessarily radiatively generated and {\it finite}.
In an $S^1/\Z_2$ orbifold description of the extra dimension, the Higgs field can be seen as a Wilson line phase on the covering circle $S^1$. From this perspective, the only gauge invariant operator that can give rise to a
Higgs potential $V(H)$ must be non-local in the extra dimension
and expressed in terms of the Wilson line $W={\cal P} \exp(i\int dy A_5)$ \cite{hos}.
Boundary local potentials for $A_5$ are forbidden by the shift symmetry (\ref{nonlinear}) \cite{vonGersdorff2}.
Being a non-local operator, $V(H)$ is finite at all orders in perturbation theory \cite{Non-local} (see also \cite{Maru:2006wa} for an explicit check up to two-loop level). Depending on the field content of the model, a radiatively induced EWSB can occur, governed by the Wilson line phase. The EWSB is thus equivalent to a Wilson line symmetry breaking.
No dependence on the UV cut-off $\Lambda$ appears in $V(H)$ and 
the hierarchy problem is solved. All GHU models are necessarily models with TeV-sized extra dimensions \cite{Antoniadis:1990ew}, since after EWSB, the $W$ mass $M_W\sim \epsilon/L$, where in natural models $\epsilon$ is a dimensionless coefficient of order ${\cal O}(10^{-1}\div 1)$.

A primordial form of the GHU idea had been advocated in refs.\cite{early} (mostly for 6D models) 
but no (semi-)realistic realization was found. The simplest GHU models one can imagine in flat 5D space,
with suitable gauge and fermion fields in the bulk giving rise to the SM zero mode spectrum, cannot work
for simple and general reasons: 
i) the Higgs, being its potential radiatively generated, is too light and ii) the top Yukawa coupling is too small.
Thanks to the advent of a more phenomenological
bottom-up approach to theories in extra dimensions, which have considerably extended 
the model building scenario, the above problems i) and ii) have now been solved.
We will show in next section two explicit and realistic models that exploit two different ideas,
an $SU(3)\times U(1)_X$ model with Lorentz symmetry breaking in the fifth dimension and
an $SO(5)\times U(1)_X$ model with large localized kinetic terms.
Before reviewing these models, in the next two subsections we show how 
to efficiently compute the Higgs effective potential and the Yukawa couplings using the holographic approach in simpler set-ups, paving the way for the more complicated situations considered in section 4.

\subsection{The one-loop Higgs effective potential}

The computation of the one-loop Higgs effective potential in GHU models provides a very good instance to appreciate the power of the holographic approach. The potential is obtained, as usual, by integrating out the whole mass spectrum of the theory in presence of a non-vanishing Higgs VEV.
In the gauge in which the field $\Sigma$ appears only at $y=0$, the bulk degrees of freedom
with $y\neq 0$ do not depend on it, the only dependence appearing through the rotation (\ref{BSigma}). 
The relevant holographic Lagrangian for the gauge fields is 
\be
{\cal L}_{quad.}^H(\Sigma) = -\frac{P^{\mu\nu}_t }{2g_5^2} \, C_\mu^{a^\prime}\Pi_g^{a^\prime b^\prime}(\Sigma) 
 C_\nu^{b^\prime} \,,
\label{LholoGExpSigma}
\ee
from which the potential for $\Sigma$ is easily computed to be\footnote{Recall  that 5D ghosts are decoupled in the unitary gauge $A_y = 0$ and the 4D ghosts associated to the Landau gauge  $C_\mu^l=0$ do not contribute to the Higgs potential.} (rotating to Euclidean momenta) 
\be
 V_g(\alpha)  =  \frac 32 \int\frac{d^4p_E}{(2\pi)^4} \log \bigg[{\rm Det}\, \Big(\Pi_g(i p_E,\Sigma)\Big) \bigg] \,.
\ee
The fermion contribution to the Higgs potential is also easily derived. The best choice to efficiently compute the Higgs potential is to retain, independently of the actual fermion b.c. at $y=0$,  
all the holographic fields inside a given multiplet with the same chirality components. 
In this way the same gauge fixing chosen in the gauge sector to rotate away $\Sigma$ for the whole bulk Lagrangian allows to also rotate $\Sigma$ away in the fermion sector. In this gauge, the holographic fields are rotated (keeping for definiteness the left-handed components) 
\be
\psi_L^I(y=0)= \Big(\Sigma^{-1}\Big)^I_J \chi_L^J\,\,,
\label{SigmaFer}
\ee
like the gauge fields in eq.(\ref{BSigma}), where $\Sigma$ in (\ref{SigmaFer}) is in the representation given by $\psi$.\footnote{The same rotation has to be performed to the Lagrange multiplier fields, so that eq.(\ref{L0lm}) is left invariant.} After solving for the Lagrange multipliers, we can set the $(-)$ components of $\chi_L$ to zero and finally obtain the holographic action   \cite{Panico:2007qd}
\be
{\cal L}(\Sigma) = (\bar \chi_L \Sigma)_I \Pi_L^I  (\Sigma^{-1} \chi_L)^I  \equiv \bar \chi_L^{i} \Pi_L^{i j}(\Sigma) \chi_L^{j}\,,
\label{VferGen}
\ee
where $\Pi_L^I = \Pi_L^\pm$,  the fermion form factors defined in eq.(\ref{LagHolFerm}), 
depending on the b.c. at $y=L$ of the corresponding fermion component, and $i$, $j$ run over the left-handed fermion components with $(+)$ b.c. at $y=0$. From eq.(\ref{VferGen}) we get
\be
 V_f(\alpha)  =  -2 \int\frac{d^4p_E}{(2\pi)^4} \log \bigg[{\rm Det}\, \Big(\Pi_L(i p_E,\Sigma)\Big) \bigg] \,,
 \label{VferGenP}
\ee
where ${\rm Det}$ refers only to the gauge indices, the spinorial ones being already considered 
and resulting in the overall factor 2. Equations (\ref{VferGen}) and (\ref{VferGenP}) allow us to see, without the need of any detailed computation, that not all b.c. gives rise to a non-vanishing contribution to the Higgs potential. When $\psi_L^I$  have the same b.c. at $y=L$ for any $I$, independently of what happens at $y=0$, the Higgs potential vanishes.  The form factors $\Pi_L^I$ do not depend on $I$, and hence the $\Sigma$ dependence trivially cancels from eq.(\ref{VferGen}): $\Sigma \Sigma^{-1} = I$. Similarly,
when $\psi_L^I$ have all the same b.c. at $y=0$, independently of what happens at $y=L$, the $\Sigma$ dependence cancels
in the determinant in eq.(\ref{VferGenP}).

Let us illustrate the above results with a simple example. In the notation of section \ref{gaugeHOLO},
we take  $G=SU(2)$, $H=H^\prime=U(1)$. The ``Higgs'' is a doublet given by $h_{\hat 1}$ and $h_{\hat 2}$ along the two broken generators $\sigma_{\hat 1,\hat 2}$, $\sigma_{i}$ being  the $2\times 2$ Pauli matrices (${\rm Tr}\,\sigma_{i} \sigma_{j} = 2 \delta_{ij}$): 
 \be
 \Sigma  =  \exp \Big[ i\sum_{\hat a=1,2}\frac{\sigma_{\hat a}  h_{\hat a}}{f_\pi} \Big] \,.
\label{Sigmasu2}
 \ee
Using eq.(\ref{BSigma}) and the unbroken $U(1)$ to align the VEV along $\sigma_2$, we have
$A_\mu^2(y=0)=0$, $A_\mu^1(y=0)=C_\mu^3\sin (2\alpha)$, $A_\mu^3(y=0)=C_\mu^3\cos (2\alpha)$,
where $\alpha\equiv \langle h_{\hat 2}\rangle /f_\pi$. Hence  
\be
{\cal L}_{quad.}^H(\alpha) =  \frac{2}{g_5^2 L f_\pi^2} \sum_{\hat a=1,2}p^2 h_{a}^2 - \frac{P_t^{\mu\nu}}{2g_5^2}C_\mu^3 \big[ 
\Pi^+_g(p) +\sin^2(2\alpha) (\Pi^-_g(p)-\Pi^+_g(p)) \big] C_\nu^3\,,
\ee
and
\bea
V_g(\alpha) \a = \a \frac 32 \int\frac{d^4p_E}{(2\pi)^4} \log \bigg[1+\sin^2(2\alpha) \frac{\Pi^-_g(i p_E)-\Pi^+_g(i p_E)}{\Pi^+_g(i p_E)}\bigg] \nn \\
\a = \a \frac 32 \int\frac{d^4p_E}{(2\pi)^4} \log \bigg[1+\frac{\sin^2(2\alpha)}{\sinh^2(Lp_E)} \bigg]   \nn \\
\a = \a - \frac{9}{64\pi^2 L^4} \sum_{n=1}^\infty \frac{1}{n^5} \big[\cos(4 n \alpha) - 1\big]\,,
\label{Vgauge}
\eea
where in the second line of eq.(\ref{Vgauge}) the explicit expressions (\ref{PiGauge}) for $\Pi_g^{\pm}$ have been used and
an irrelevant (divergent) $\alpha$-independent term has been added so that $V_g(0)=0$.

Let us also consider in detail the contribution of a massive fermion doublet $\psi^I$ ($I=1,2$) for the same
symmetry breaking pattern $SU(2)\rightarrow U(1)$ considered above, 
for all possible choices of b.c. for $\psi$. These are in total 16, but only 8 are independent,  the other half being simply obtained by
an exchange of chirality $L \leftrightarrow R$. Let us choose $\psi^{1,2}_L$ as holographic fields and take $(+)$ b.c.
for $\psi^1_L$. As we mentioned, a non-trivial contribution to the potential arises when $\psi_L^1$ and $\psi_L^2$ have opposite b.c. 
at both $y=0$ and at $y=L$. This fixes $\psi_L^2$ to be $(-)$ at $y=0$ and from eq.(\ref{SigmaFer}) we have $\psi_L^1 = \cos (\alpha) \,\chi_L^{1}$, $\psi_L^2 = \sin (\alpha) \,\chi_L^{1}$. We are left with two options of b.c. at $y=L$, namely i) $\psi_R^1(L)=0$ or ii) $\psi_L^1(L)=0$. In the two cases, we have  i) ${\rm Det}\, \Pi_L(ip_E,\Sigma) = \cos^2(\alpha) \Pi_L^+(ip_E)+ \sin^2(\alpha) \Pi_L^-(ip_E)$, ii) ${\rm Det}\, \Pi_L(ip_E,\Sigma) = \cos^2(\alpha) \Pi_L^-(ip_E,\Sigma) + \sin^2(\alpha)  \Pi_L^+(ip_E)$. The Higgs potential, shifted so that $V_f(0)=0$, is then 
\bea
i) V_f(\alpha) \a = \a -2\! \int\!\!\frac{d^4p_E}{(2\pi)^4} \log \bigg[1+\frac{(m^2+p_E^2)\sin^2(\alpha) }{p_E^2\sinh^2(L\sqrt{p_E^2+m^2})} \bigg]   ,  \nn \\
ii) V_f(\alpha) \a = \a -2\! \int\!\!\frac{d^4p_E}{(2\pi)^4} \log \bigg[1-\frac{(m^2+p_E^2)\sin^2(\alpha)}{m^2+p_E^2 \cosh^2(L\sqrt{p_E^2+m^2})} \bigg]  \,. 
\label{Vfermion}
\eea
The integrals do not seem to admit  simple analytic expressions for generic $m$. When $m=0$, they simplify to 
\bea
 i) V_f(\alpha)_{m\rightarrow 0} \a = \a   \frac{3}{16\pi^2 L^4} \sum_{n=1}^\infty \frac{1}{n^5} \big[\cos(2 n \alpha) - 1\big],  \\
 ii) V_f(\alpha)_{m\rightarrow 0} \a = \a \frac{3}{16\pi^2 L^4} \sum_{n=1}^\infty \frac{(-1)^n}{n^5} \big[\cos(2 n \alpha) - 1\big] \,. \nn 
\label{Vfermionexp}
\eea
It is straightforward to explicitly check that $V_f(\alpha)$ vanishes for the remaining 6 choices of boundary conditions. 
In particular, when $\psi_L^2$ is $(+)$ at $y=0$, so that $\Pi_L$ is a $2\times 2$ matrix, $|{\rm Det}\, \Pi_L| = \Pi_L^+ \Pi_L^-$ does not depend on $\alpha$. The potentials $V_g(\alpha)$ and $V_f(\alpha)$ agree with the expressions found with a more direct, but laborious, KK approach (see e.g. \cite{Scrucca:2003ra}) and are manifestly finite, as expected.

Notice that if we choose different chirality components as holographic fields
among fermions in the same multiplet, the fermion contribution to the potential will {\it not} be given only by the holographic Lagrangian,
since the bulk fields with $y\neq 0$ would have a $\Sigma$-dependent mass spectrum that should be taken into account.
Needless to say, taking into account the bulk contribution as well, the ending result would be the same, but with a more laborious procedure, that spoils the utility of the holographic approach.

The total Higgs potential at one-loop level is the sum over the gauge and fermion field contributions: $V(\alpha) = V_g(\alpha) + V_f(\alpha)$.  Due to the exponential suppression, for large $p_E$,  of the form factors  appearing inside the logarithm in $V_g(\alpha)$ and $V_f(\alpha)$, and the power suppression in $p_E$ due to phase space at low momentum, the main contribution to the momentum integration in the potential is given for $p_E L \sim 1$.
For such values of $p_E$, the form factors inside the logarithm are smaller than one, and hence it is reasonable to expand the log and keep the leading term. In this way, the total potential is simply 
\be
L^4 V_{app}(\alpha) = c  \sin^2 (2\alpha) - d \sin^2 (\alpha)\,,
\label{PotApprox}
\ee
where $c>0$ and $d$ are easily derived from the explicit forms (\ref{Vgauge}), (\ref{Vfermion}).
The potential (\ref{PotApprox}) has extrema at $\alpha_0=0,\pi/2$. If $|d/c|<4$, an other extremum is
at $\cos (2\alpha_0) = d/(4c)$.  When $\alpha_0\neq 0$,  a gauge symmetry breaking is induced.  For any value of $|d/c|<4$, the latter extremum is always a maximum and hence the only non-trivial minimum is $\alpha_0=\pi/2$.
The  ``$W$" and ``Higgs" masses in this toy model are easily computed.
From the second line in eq.(\ref{Vgauge}), one can directly read the mass of the W as the first mass state with $p_E^2=- m_W^2$:
\be
M_W  = \frac{2\alpha_0}{L}\,.
\label{mW}
\ee
The Higgs mass squared is given by 
\be
M_H =\frac{\sqrt{V^{\prime\prime}(\alpha_0)}}{f_\pi}=M_W L^2 \frac{g_4}{4 \alpha_0}\sqrt{V^{\prime\prime}(\alpha_0)} \simeq
M_W\frac{g_4(d+4c)}{\sqrt{2}\pi}
\,.
\label{HiggsMass}
\ee
In eq.(\ref{HiggsMass}), $\alpha_0=\pi/2$, 
$g_4(\alpha_0)$ is the 4D gauge coupling constant defined as in eq.(\ref{defggprime}): 
\be
g_4^2(\alpha_0)=\frac{3g_5^2}{L[2+\cos (4\alpha_0)]}\,,
\ee
and the Higgs field has been canonically normalized by setting $f_\pi^2 = 4/(g_5^2 L)$.
Since generally $d,c\lesssim 1$, the Higgs tends to be too light.
We can give a better estimate of the Higgs mass by taking the specific example of one fermion multiplet in the case i), namely $\psi_L$ $(+)$ at $y=L$ and with $m=0$, in which case one has
\be
V^{\prime\prime}(\alpha_0=\pi/2) = \frac{45 \zeta(3)}{16\pi^2L^4}
\ee
giving 
\be
\frac{M_H}{M_W}\approx 0.1 g_4\,.
\ee
The smallness of the Higgs mass is essentially due to the radiative nature of its potential, resulting in a too small quartic coupling. Values of $\alpha_0$ of order 1 give also rise, by means of eq.(\ref{mW}), to too low compactification scales. 
In order to get $\alpha_0\ll 1$, and hence reasonable compactification scales,
it is enough to engineer a model where the potential looks like
\be
L^4 V_{app}(\alpha) = c\sin^2(\alpha) -d \sin^2(2\alpha)\,,
\label{PotApprox2}
\ee
so that the non-trivial extremum  at $\cos (2\alpha_0) = c/(4d)$ for $|c/d|<4$, is now a minimum for $d>0$.
The $\alpha$ factors entering in the fermion and gauge contribution to the potential are determined by
group theory, so with a proper choice of gauge groups and fermion representations it is not difficult to get potentials
like (\ref{PotApprox2}).  When $|c/d|$ is just slightly below the value $4$, $\alpha_0\ll 1$.
This requires a fine-tuning,  unless a natural mechanism is at work, favouring $|c/d|\simeq 4$ among other possible values. Its amount can be estimated by adapting the well-known Barbieri-Giudice relation \cite{Barbieri:1987fn}
to our situation:
\be
f =  \sqrt{\sum_i \left( \frac{\partial\log \alpha}{\partial\log k_i}\right)^2}\,,
\label{finetuning}
\ee
where $k_i$ are the microscopic input parameters (such as the bulk fermion masses) from which $V(\alpha)$ depends on. 
Evaluating eq.(\ref{finetuning})  at  $\alpha_0$ gives
\be
f \simeq \frac{\cot (2\alpha_0)}{2\alpha_0}   \sqrt{\sum_i \left( \frac{\partial\log c/(4d)}{\partial\log k_i}\right)^2}\sim
\frac{1}{4\alpha_0^2}\,,
\label{finetuning2}
\ee
where in the last expression we have expanded for small $\alpha_0$ and neglected possible contributions coming
from the square root factor. Tentatively, and considering that fine-tuning issues should always be taken with some grain of salt, eq.(\ref{finetuning2}) allows us to conclude that values of $\alpha_0= {\cal O}(10^{-1})$ are moderately tuned and can be considered acceptable, whereas $\alpha_0= {\cal O}(10^{-2})$ or smaller cannot be seen as a satisfactory solution to the little hierarchy problem.
In the case in which the $\alpha$ periodicity of the fermion and gauge contribution is the same, so that $V_{app}(\alpha) \propto \sin^2 \alpha$, the only extrema are at $\alpha_0=0$ and $\alpha_0=\pi/2$.

\subsection{Yukawa couplings}

The Yukawa couplings are also readily derived holographically. Instead of solving for the Lagrange multiplier fermions $\lambda_R$, setting the $(-)$ components of $\chi_L$ to zero, as tacitly done in 
deriving the Lagrangian (\ref{VferGen}), we now keep the $\lambda_R$, so that 
\be
{\cal L}(\chi_L,\lambda_R,\Sigma) = (\bar \chi_L \Sigma)_I \Pi_L^I  (\Sigma^{-1} \chi_L)^I  +(\bar \lambda_R^{a^\prime} \chi_L^{a^\prime} + h.c.)\,,
\label{VferYuk}
\ee
with $a^\prime\in {\cal G}/{\cal H}^\prime$, and instead  solve for the $(-)$ components of $\chi_L$.
In this way, the holographic Lagrangian for the low-energy fermion excitations is expressed in terms of the $(+)$ components of $\chi_L$ and of the now dynamical Lagrange multipliers $\lambda_R$.
 In order to illustrate the procedure, consider the same $SU(2)\rightarrow U(1)$ toy model analyzed in subsection 3.1, taking the b.c. i), the only ones giving rise to chiral zero modes. Solving for $\chi_L^2$ gives
\be
\chi_L^2 = \Pi^{-1}(\alpha)\Big[ \sin(\alpha)\cos(\alpha) (\Pi_L^+-\Pi_L^-)\chi_L^1 -\frac{\pslash}{p}  \lambda_R\Big] \,,
\label{chiL2Expl}
\ee
where $\Pi(\alpha) = \sin^2(\alpha) \Pi_L^++\cos^2(\alpha) \Pi_L^-$.
Plugging eq.(\ref{chiL2Expl}) back in eq.(\ref{VferYuk}) gives
\be
{\cal L}(\chi_L^1,\lambda_R,\alpha) = \Pi^{-1}(\alpha)\bigg[ \bar \chi^1_L \frac{\pslash}{p} \Pi_L^+ \Pi_L^- \chi^1_L - \bar \lambda_R \frac{\pslash}{p} \lambda_R + \cos(\alpha)\sin(\alpha) ( \Pi_L^+- \Pi_L^-) (\bar \chi^1_L \lambda_R + \bar \lambda_R \chi^1_L) \bigg]\,.
\label{LholoYukExp}
\ee
For $\alpha=0$, $\chi_L^1$ and $\lambda_R$ are decoupled, and each gives rise to a massless zero mode. When $\alpha\neq 0$,
the two mode towers are coupled and the mass eigenvalues $M_n^2$ given by the zeros of the determinant of the $2\times 2$ kinetic term. Explicitly, one has
\be
M_n^2\Big[ \cos\big(2L \sqrt{M_n^2-m^2}\big) -\cos(2\alpha) \Big] = m^2 \big[1-\cos(2\alpha)\big]\,.
\label{Mn}
\ee
We can expand in powers of the momentum the holographic Lagrangian (\ref{LholoYukExp}) to get the low energy
effective theory.  At leading order, one has
\be
{\cal L}(\chi_L^1,\lambda_R,\alpha) = Z_L \bar \chi_L^1 \,\pslash \,\chi_L^1 +  Z_R \bar \lambda_R \,\pslash\, \lambda_R - \tan(\alpha)  (\bar \chi^1_L \lambda_R + \bar \lambda_R \chi^1_L) \,,
\ee
with
\be
Z_L = \frac{e^{-Lm}\sinh(Lm) }{\cos^2(\alpha)m}\,, \ \ \ \ 
Z_R = \frac{e^{Lm}\sinh(Lm) }{\cos^2(\alpha)m}\,.
\label{ZLZR}
\ee
Rescaling the fermions $\chi_L^1\rightarrow \chi_L^1/\sqrt{Z_L}$, and $\lambda_R\rightarrow \lambda_R/\sqrt{Z_R}$ to
canonically normalized fields, and expanding at linear order in $\alpha$ (assumed to be $\ll 1$), we can finally read off the induced low-energy Yukawa coupling \cite{Burdman:2002se}
\be
|Y(m)|=  \frac{\sqrt{g_5^2 L} m}{\sqrt{2}\sinh(Lm)}\simeq  \frac{g_4}{\sqrt{2}}\frac{L m}{\sinh(Lm)}\,.
\label{Yuk}
\ee
For $m=0$, $Y(0) = g_4/\sqrt{2}$ and no hierarchical Yukawa's are possible. Allowing a non-vanishing $m$, however,
not only solves the problem but also gives rise in a natural way to exponentially suppressed Yukawa's.
All Yukawa couplings can be nicely accommodated in this way, with the exception of the top quark.

Hierarchical Yukawa couplings are also obtained by introducing from the beginning localized chiral fermions, say at $y=0$.
Localized fermions transform only under the group $H^\prime$ and hence no direct couplings between them and the Higgs is allowed.
The only way to generate a Yukawa coupling is by mixing them with massive bulk fermion fields. If no other chiral field
is necessary from the bulk, one can introduce a pair of fermion fields $\psi$ and $\tilde \psi$, with opposite b.c. and bulk mass terms 
$m(\bar{\tilde\psi} \psi +\bar \psi \tilde \psi)$, so that no zero modes will be generated.\footnote{From an orbifold perspective, the choice of introducing the bulk mass term $M(\bar{\tilde\psi} \psi +\bar \psi \tilde \psi)$ and no other, is dictated by the orbifold parity, being the only one even under the orbifold projection.}
Such possibility has been advocated in \cite{CGM} for GHU models in 6D and used in \cite{Scrucca:2003ra,Panico:2005dh,Panico:2006em} for GHU models in 5D.
It is less economical than the former, but it allows more flexibility, since now the Yukawa couplings
also depends on the boundary-bulk mixing mass terms. The Yukawa's so generated are always smaller than (\ref{Yuk}),
recovering eq.(\ref{Yuk}) in the limit of infinite boundary-bulk mixing mass terms.
The problem of the top Yukawa coupling still persists.
The Yukawa coupling (\ref{Yuk}) depends on the gauge group representation of
the bulk fermion under the group $G$, and Clebsch-Gordan like coefficients can appear in eq.(\ref{Yuk}). 
Choosing fermions in representations with high enough rank allow to accommodate the top quark \cite{Cacciapaglia:2005da}, although care has to be paid with high rank fermions \cite{Scrucca:2003ra}, since they lower the range of validity of the 5D theory, as estimated by  Na\"{\i}ve Dimensional Analysis (NDA)  \cite{NDA}.

Interestingly enough, both the mass eigenvalues (\ref{Mn}) and the low-energy Yukawa coupling (\ref{Yuk}) are already
encoded in the fermion contribution to the Higgs effective potential (\ref{Vfermion}).
Indeed, recalling that the one-loop Casimir energy given by a 4D fermion of mass $M$ is
$V=-2\int d^4p_E/(2\pi)^4 \log(p^2_E+M^2)$, the mass eigenvalues (\ref{Mn}) are easily obtained by 
setting to zero the argument of the logarithm in (\ref{Vfermion}) and taking $p_E^2=-M_n^2$.\footnote{Being
careful in distinguishing a zero from a pole. This is done by looking at the sign of the residue.}
Similarly, by expanding up to quadratic order in $p_E^2$ and in $\alpha^2$, one easily recovers eq.(\ref{Yuk}).

\section{Model Building}

In this section we finally build realistic models.
The minimal gauge group extensions of the electroweak SM group giving rise to
pseudo-Goldstone bosons with the SM Higgs quantum numbers and nothing more, are $SU(3)$ and $SO(5)$.
In both cases, extra $U(1)$ factors are also needed to get the correct weak-mixing angle.
As we have reviewed in section 3, GHU models in flat space have to face the quantitative
problem of getting a sufficiently heavy Higgs, top and compactification scale, so that some new qualitative ingredients have to be added
to the minimal toy models studied in section 3. We will consider in the next two subsections two possible extensions that allow
to get realistic models. The first, based on an $SU(3)$ model,  advocates an explicit tree--level breaking of the Lorentz {\rm SO(4,1)} symmetry \cite{Panico:2005dh}, so that the Yukawa coupling is not tied to the gauge coupling as in (\ref{Yuk}), but can be bigger. In this way, the top and Higgs mass problems are solved and, with a modest fine-tuning, the compactification scale is also above current experimental bounds. The second is based on an $SO(5)$ model, where large localized gauge kinetic terms are introduced. 
Models based on the group $SO(5)$ are more promising, since they have an automatic custodial protection that suppresses
otherwise large corrections to the $\hat T$ parameter \cite{Agashe:2003zs}. Large localized gauge kinetic terms were already advocated in \cite{Scrucca:2003ra} but for the $SU(3)$ model where, in absence of a custodial symmetry, lead to too large values for $\hat T$. $SO(5)$ models with large localized gauge kinetic terms might also be seen as a
useful way to construct effective composite Higgs models at the TeV scale.

We review in subsection 4.1 the construction of the $SU(3)$ model, and presents the $SO(5)$ model in subsection 4.2.
The $SO(5)$ model has actually never been considered in flat space, so that the results appearing in 4.2 are new, 
although the model we consider is the flat space version of a model already considered in warped space \cite{Contino:2006qr}.

\subsection{$SU(3)\times U(1)\times U(1)^\prime$ model}

The minimal gauge group implementing the GHU idea is $SU(3)$. As mentioned, the group $SU(3)$ alone will
give rise to the wrong weak-mixing angle $\sin^2\theta_W = 3/4$, so that at least an extra $U(1)$ has to be added \cite{Scrucca:2003ra}.

A potentially realistic $SU(3)_w$ model with gauge-Higgs unification in flat space can be obtained
by advocating an explicit tree--level breaking of the Lorentz {\rm SO(4,1)} symmetry \cite{Panico:2005dh}.
Indeed, the smallness of the top Yukawa coupling is a consequence of the $SO(4,1)$ Lorentz symmetry,
linking the Yukawa to the gauge coupling, as in eq.(\ref{Yuk}).  Breaking the SO(4,1)/SO(3,1) symmetry
(so that the usual SO(3,1) Lorentz symmetry is unbroken) give a way to increase the couplings between the Higgs field 
and the fermions in a 5D gauge-invariant way. 
The 5D model we review, closely following \cite{Panico:2006em}, is essentially the Lorentz breaking version of the minimal $SU(3)_w$ model proposed in \cite{Scrucca:2003ra}, where a further $\Z_2$ ``mirror'' symmetry is added. 
The $\Z_2$ symmetry, motivated by naturalness arguments, essentially consists in doubling
a subset of bulk fields $\phi$ in pairs $\phi_1$ and $\phi_2$ and requiring a symmetry under the
interchange $\phi_1\leftrightarrow \phi_2$. 

The b.c. of all the fields in this model are the standard ones
coming from an orbifold projection, so it will be useful to adopt in the following the orbifold perspective. 
The gauge group is taken to be of the form $G \times G_1 \times G_2$,
with  $G=SU(3)_w\times SU(3)_c$ and $G_i = U(1)_i$, although other choices are allowed.
The  $\Z_2$ orbifold projection is embedded non-trivially in the electroweak $SU(3)_w$
group only, by means of the matrix
\begin{equation}
P = e^{2i\pi   t_3} =
\left(
\begin{matrix}
-1 \a 0 \a 0 \cr
 0 \a -1\;\;\, \a 0 \cr
 0 \a  0 \a 1 \cr\end{matrix}\;
\right)\;,\label{Rtwist1}
\end{equation}
where $t_a$ are the $SU(3)$ generators,
normalized as ${\rm Tr}\, t_a t_b = \delta_{ab}/2$.\footnote{The conventions and notation used here  do not coincide with those taken in \cite{Panico:2006em}, but have been changed  in order to keep them the same throughout the paper.}
The abelian $U(1)_i$ fields satisfy the following b.c. (omitting for simplicity vector indices):
\be
A_1(y\pm 2\pi R) =  A_2(y) \,,
\;\;\;\;\;\;\;A_{1}(-y) =  \eta A_{2}(y) \,,
\label{bound-cond}
\ee
where $\eta_\mu=1$, $\eta_5=-1$.
The unbroken gauge group at $y=L$ is $H=SU(3)_c \times SU(2)\times U(1) \times U(1)_1 \times U(1)_2$,
while at $y=0$ we have $H^\prime=SU(3)_c\times SU(2) \times U(1) \times U(1)_+$,
where $U(1)_+$ is the diagonal subgroup of $U(1)_1$ and $U(1)_2$. 
Under the mirror symmetry, the linear combinations $A_{\pm}=(A_1\pm A_2)/\sqrt{2}\rightarrow \pm A_{\pm}$, so we can assign a
multiplicative charge +1 to $A_+$ and $-1$ to $A_-$.
The massless 4D gauge fields are the vector bosons  in the adjoint of $SU(2)\times U(1)\subset SU(3)_w$, the $U(1)_+$ and the gluon gauge fields $A_c$. The $SU(3)_{c}$ and $SU(2)$ gauge groups are identified 
with the SM $SU(3)$ and $SU(2)$ factors, while the hypercharge $U(1)_Y$ is
the diagonal subgroup of $U(1)$ and $U(1)_+$.
The Higgs field arises  from the zero mode $A_{w}^{4,5,6,7}$
components of the $SU(3)_w$ gauge fields. In the holographic gauge-fixing of subsection 2.2 it can be written as
\be
\Sigma = \exp\Big[ i \sum_{\hat a=1}^4  \frac{2 t^{\hat a+3} h^{\hat a}}{f_\pi}  \Big]\,, \ \ \ 
f_\pi = \frac{2}{g_5 \sqrt{L}}\,,
\ee
where $g_5$ is the $5D$ charge of the $SU(3)_w$ group.
The extra $U(1)_X$ gauge symmetry which survives the orbifold
projection is anomalous and its  gauge boson gets a mass of the order of the cut-off scale $\Lambda$ of the model.
  
A certain number of couples of bulk fermions $(\Psi_1,\widetilde \Psi_1)$
and $(\Psi_2,\widetilde \Psi_2)$ are introduced, with identical quantum numbers under the group $G$
and opposite orbifold parities. The couples $(\Psi_1,\widetilde \Psi_1)$ are charged under $G_1$ and neutral under $G_2$ and, by mirror symmetry, the same number of couples $(\Psi_2,\widetilde \Psi_2)$ are charged under $G_2$ and neutral under $G_1$.
No bulk field is simultaneously charged under both $G_1$ and $G_2$.
In total, for each SM generation, one pair of
couples $(\Psi_{1,2}^{u},\widetilde \Psi_{1,2}^{u})$ in the ${\bar{\bf 3}}$ 
of $SU(3)_w$ and one pair of couples $(\Psi_{1,2}^{d},\widetilde \Psi_{1,2}^{d})$ in the ${\bf 6}$ 
of $SU(3)_w$ are introduced. Both pairs have $U(1)_{1,2}$ charge +1/3 and are in the {\bf 3} 
of $SU(3)_{s}$. The b.c. of these fermions  follow from the twist matrix 
(\ref{Rtwist1}) and eqs.~(\ref{bound-cond}).
 Massless chiral fermions with charge +1 with respect
to the mirror symmetry, localized at $y=0$, are also introduced.
As  far as EWSB is concerned, we can focus on the
top and bottom quark only, neglecting all the other SM matter fields, which can be accommodated.
Mirror symmetry and the b.c. (\ref{bound-cond})
imply that the localized fields can couple only to $A_+$. Hence, we have an $SU(2)$
doublet $Q_L$ and two singlets $t_R$ and $b_R$, all in the ${\bf 3}$ 
of $SU(3)_{s}$ and with charge $+1/3$ with respect to the $U(1)_+$ gauge field $A_+$.

The most general 5D Lorentz breaking effective Lagrangian density, gauge invariant and mirror symmetric,
up to dimension $d<6$ operators, is:\footnote{Strictly speaking, 
the Lagrangian (\ref{fullL}) is not the most general one, since we are neglecting
all bulk terms which are odd under the $y\rightarrow -y$ parity transformation and
can be introduced if multiplied by odd couplings. If not introduced, such couplings are not generated
and thus can consistently be ignored.}
\be
\mathcal{L}= \mathcal{L}_g+ \mathcal{L}_\Psi+2\delta(y) \mathcal{L}_0+2\delta(y-L) \widehat{\mathcal{L}}_L\,,
\label{fullL}
\ee
with
\bea
\mathcal{L}_g= \a\a \sum_{i=1,2}\bigg[-\frac{1}{4}F_{i\mu\nu}
F^{i\mu\nu} -\frac{\rho^{2}}{2}F_{i\mu y} F^{i\mu y} \bigg]   -\frac{\epsilon}{4}F_{1\mu\nu}
F^{2\mu\nu}-\frac{\tilde \rho^{2}}{2}F_{1\mu y} F^{2\mu y}
\nn \\
\a\a
-\frac{1}{2}{\rm Tr}\,F_{\mu\nu}F^{\mu\nu} -
\rho_w^2{\rm Tr}\,F_{\mu y}F^{\mu y} -\frac{1}{2}{\rm Tr}\,G_{\mu\nu}G^{\mu\nu} -
\rho_s^2{\rm Tr}\,G_{\mu y}G^{\mu y} \,, \label{Lgauge} \\
\mathcal{L}_\Psi = \a\a \sum_{i=1,2}
\sum_{a=t,b} \bigg\{\overline\Psi_i^{a} \big[i\Dslash_{4}(A_i)+k_a D_5(A_i)
\gamma^5 \big] \Psi_i^{a} \label{LPsi} \\
\a\a \ \ \ \
+\overline{\widetilde\Psi}_i^{a} \big[i\Dslash_{4}(A_i)+\widetilde k_a D_5(A_i) \gamma^5
\big] \widetilde\Psi_i^{a} -
m_a (\overline{\widetilde\Psi}_i^{a} \Psi_i^{a} +\overline{\Psi}_i^{a} \widetilde\Psi_i^{a})\bigg\} \,, \nn \\
\mathcal{L}_0= \a\a \overline Q_L i \Dslash_4(A_+) Q_L +
\overline t_R i \Dslash_4(A_+)t_R + \overline b_R i \Dslash_4(A_+) b_R
\nn \\
\a\a + \big(e_1^t \overline Q_L \Psi_{+}^{t} + e_1^b \overline Q_L \Psi_{+}^{b}
 + e_2^t \overline t_R \Psi_{+}^{t} + e_2^b \overline b_R \Psi_{+}^{b}+\mathrm{h.c.} \big)+
\widehat{\mathcal{L}}_0 \label{Lloc0}\,.
\label{LlocPi}
\eea
In eq.~(\ref{Lgauge}), we have denoted by $G=DA_{c}$ the $SU(3)_{c}$ field strength, for simplicity
we have only schematically written the dependencies of the covariant derivatives on the gauge fields
and we have not distinguished the doublet and singlet
components of the bulk fermions in eq.~(\ref{Lloc0}), denoting all of them simply as $\Psi^{t}_+$ and
$\Psi^{b}_+$. 
Extra brane operators, such as for instance localized kinetic terms,
are included in $\widehat{\mathcal{L}}_0$ and $\widehat{\mathcal{L}}_L$.
Additional Lorentz violating bulk operators like $\overline\Psi \gamma^5 \widetilde\Psi$,
$\overline\Psi \partial_y \Psi$ or $\overline\Psi i\Dslash_4 \gamma^5 \Psi$ can be forbidden by requiring
 invariance under the inversion of all spatial (including the compact one) coordinates,
under which any fermion transforms as $\Psi\rightarrow \gamma^0 \Psi$. 
This $\Z_2$ symmetry is a remnant of the broken $SO(4,1)/SO(3,1)$ Lorentz generators.

The mirror symmetry constrains the Lorentz violating factors for periodic and antiperiodic fermions
to be the same: $k_+=k_-\equiv k$, $\widetilde k_+=\widetilde k_-\equiv \widetilde k$ for both the
${\bf \overline 3}$ and ${\bf 6}$ representations, resulting in a significant reduction of the fine-tuning in the model.
All SM fields are even under the mirror symmetry, implying that the lightest $\Z_2$ odd 
state in the model is absolutely stable. In a (large) fraction of the parameter space of the model such state 
is the first KK mode of the $A_-$ gauge field and it has in fact been shown to be a viable DM candidate  \cite{Regis:2006hc}.

A detailed study of the model using the general Lagrangian (\ref{fullL}) is a too complicated task.
In order to simplify our analysis, we take $\epsilon = \tilde \rho^2 = 0$,\footnote{This assumption was implicit in \cite{Panico:2006em}, where these terms had not been included in eq.(\ref{Lgauge}).} $k_a=\widetilde k_a$ and set $\rho_w=1$.
The latter choice can always be performed without loss of generality by rescaling the compact
coordinate, and hence the radius of compactification as well as the other parameters of the theory.
We also neglect all the localized operators which are encoded in
$\widehat{\mathcal{L}}_0$ and $\widehat{\mathcal{L}}_\pi$.

The $W$ boson mass is given by 
\be
M_W=\frac{\alpha}{L}\,,
\label{mwSU3}
\ee
where
\be
\alpha \equiv  \frac{\langle h \rangle }{f_\pi}\,, \ \ \ h=\sqrt{\sum_{\hat a=1}^4 h_{\hat a}^2} \,.
\label{alphaDef}
\ee
The top Yukawa coupling reads, for large mixing mass parameters, 
\be
Y_{t} \simeq  k_t g_4 \frac{2 m_t L /k_t}{\sinh{(2m_t L /k_t)}}\,,
\label{mtop-app}
\ee
and shows the effect of the Lorentz breaking parameter $k_t$. 
For $k_t=1$ one recovers the Lorentz invariant situation and a Yukawa coupling of the
form (\ref{Yuk}). When $k_t>1$, $Y_t>g_4$ and for $k_t\sim 2\div 3$ the top mass
in the correct range is recovered. Notice that $k_t$ is essentially the only Lorentz symmetry
breaking parameter that we really need, all the other ones having being introduced 
for consistency and naturalness. The lightest non-SM particles are colored fermions with a mass of order the bulk mass parameter
$M_b$. Before EWSB they are given by an $SU(2)$ triplet with hypercharge $Y=2/3$, a doublet
with $Y=-1/6$ and a singlet with $Y=-1/3$. For the typical values of the parameters
needed to get a realistic model, the mass of these states is of order $1-2\ \rm TeV$.

The computation of the one-loop Higgs effective potential associated to the Lagrangian (\ref{fullL})
is a bit involved, but it is conceptually straightforward, using the techniques introduced in the previous sections.
The full Higgs effective potential is dominated by the fermion contribution.
The presence of bulk antiperiodic fermions, whose coupling with the Higgs
are the same as for periodic fermions due to the mirror symmetry, allows for a natural
partial cancellation of the leading Higgs mass terms in the potential,
then lowering the position of its global minimum $\alpha_{0}$.
{}The physical Higgs mass reads
\begin{equation}
M_H =\frac{\sqrt{V^{\prime\prime}(\alpha_0)}}{f_\pi}=
\displaystyle \frac{g_4 L}{2} \,\sqrt{V^{\prime\prime}(\alpha_0)}\,.
\label{MH}
\end{equation}
The leading fermion contribution to $V(\alpha)$ is proportional to $k_t^4$, so that
the latter cures at the same time the problem of a too light top and Higgs fields. 

It turns out that the 4 most constrained flavour and CP conserving dimension 6 operators in this model are 
those associated to the universal parameters $\widehat S$, $\widehat T$, $W$ and $Y$ introduced  before
(see  \cite{Panico:2005dh} for an order of magnitude estimate of the bounds arising from the calculable FCNC effects).
All light fermions are almost completely localized at $y=0$ and their couplings with the SM
gauge fields are universal and not significantly distorted. Even the $Z b\bar b$ coupling deviation
is sub-leading with respect to $\widehat S$, $\widehat T$, $W$ and $Y$.
Using eqs.(\ref{STWYdef}), one finds, at tree--level and at leading order in $\alpha=M_W L$, 
\be
\widehat S =\displaystyle\frac23 M_W^2 L^2, \ \ 
\widehat T = M_W ^2L^2, \ \ 
Y =\displaystyle\frac{\rho^2\sin^2(\theta_W)+1+2\cos(2\theta_W)}{
9\rho^2\cos^2(\theta_W)}M_W^2 L^2,  \ \  
W = \displaystyle\frac13M_W^2 L^2.
\label{obliquepar}
\ee
The lower bound on the compactification scale that one gets by a $\chi^2$ fit using the values in eq.~(\ref{obliquepar}) is
$1/L \gtrsim 1.3-1.6\ \rm TeV$,
which corresponds to $\alpha_0 \lesssim 1/20$. According to eq.(\ref{finetuning2}), the fine-tuning associated to such values of $\alpha_0$
is $\simeq 1\%$, in agreement with more accurate estimates performed in \cite{Panico:2006em}. Using the more refined definition of fine-tuning as given in \cite{Anderson:1994dz}, which takes into account for the possible presence of a generic sensitivity, it has been pointed out in \cite{Panico:2006em} that the intrinsic tuning to get $\alpha_0 \lesssim 1/20$ is reduced to $\sim 10\%$.

The Lorentz violating factors affects the range of perturbative validity of the 5D effective theory, as estimated using NDA. 
For $k_t\leq 3$, the cut-off of the model is estimated to be $\Lambda\geq 10/L$, ensuring a large enough perturbative range.

\subsection{$SO(5)\times U(1)_X$ model}

The model we analyze below is the flat space version of one of the models considered in \cite{Contino:2006qr}
and denoted there MCHM$_5$.  The bulk gauge group is $G=SU(3)_c\times SO(5)\times U(1)_X$. We denote by $g_5$ and $g_{5X}$ the
5D gauge coupling constant of $SO(5)$ and $U(1)_X$, respectively.
The unbroken group at $y=L$ is $H=SU(3)_\times SO(4)\times U(1)_X\simeq SU(3)_c\times SU(2)_L \times SU(2)_R \times U(1)_X$. The unbroken group at $y=0$ is $H^\prime=SU(3)_c\times SU(2)_L\times U(1)_Y=G_{SM}$, where $Y=X+T_{3R}$. 
Neglecting from now on the color $SU(3)_c$ factor,  the b.c. for the (non-canonically normalized) gauge fields are as follows:
\bea
F_{\mu y,L}^a \a = \a F_{\mu y,R}^a = F_{\mu y,X}= 0\,, \ \ A^{\hat a}_\mu=0\,,\ \ \ a=1,2,3\,, \hat a \in {\cal G}/{\cal H}\,, \hspace{1.1cm}  y=L,  \\
F_{\mu y,L}^a \a = \a F_{\mu y,R}^3+F_{\mu y,X} =0, \ \ 
A^{\hat a}_\mu=A^{1,2}_{\mu,R}= 0, \ \  A_{\mu,R}^3=A_{\mu,X} =B_\mu\,, \ \ y=0. \nn 
\eea
We introduce localized gauge kinetic terms at $y=0$. 
The EW gauge Lagrangian is
 \be
{\cal L}_g = {\cal L}_{5g}+{\cal L}_{4g,0},
\ee
with
\bea
{\cal L}_{5g} \a=\a \int_0^L \!\! dy\bigg\{ \frac{1}{2g_5^2}  {\rm Tr} \Big[-\frac 12 F_{\mu\nu}^2 + (\partial_y A_\mu)^2   \Big]+ \frac{1}{2g_{5X}^2}   \Big[-\frac 12 F_{\mu\nu,X}^2 + (\partial_y A_{\mu,X})^2  \Big]\bigg\} .\nn \\
{\cal L}_{4g,0} \a=\a -\frac{\theta L}{4g_5^2}  \sum_{a=1}^3 (W_{\mu\nu}^a)^2 - \frac{\theta^\prime L}{4g_{5X}^2}  B_{\mu\nu}^2\,,
\eea
$\theta$ and $\theta^\prime$ dimensionless parameters and the generators normalized as ${\rm Tr}\, t_a t_b=\delta_{ab}$ in the fundamental representation.\footnote{The different choice of normalization of the generators in the $SU(3)$ and $SO(5)$ group is due to the different embedding of $SU(2)_L$ in the two cases.} The Higgs field is given by the $G/H$ components of $A_y$ (see Appendix C for our choice of $SO(5)$ generators):
\be
\Sigma = \exp\Big[\sum_{\hat a=1}^{4} i\frac{ \sqrt{2} t^{\hat a} h_{\hat a} }{f_\pi}\Big]\,, \ \ \ f_\pi=\frac{\sqrt{2}}{g_5\sqrt{L}}\,.
\label{Sigmaso5}
\ee
The holographic Lagrangian for the SM gauge fields $W_\mu^a$ and $B_\mu$ is easily derived. In terms of the form factors defined in eq.(\ref{LagPi}), we get
\bea
\Pi_{W_aW_b} \a =\a  \frac{\delta_{ab}}{2g_5^2} \Big[2\Pi^+_g + s_\alpha^2\big(\Pi^-_g -\Pi^+_g\big) + 2p^2\theta L  \Big] \,, \nn \\
\Pi_{W_3 B} \a = \a  \frac{1}{2g_5^2}  s_\alpha^2 (\Pi^+_g -\Pi^-_g) \,
\label{Piso5} , \\
\Pi_{B B} \a = \a  \frac{1}{g_{5X}^2}  (\Pi^+_g +p^2 \theta^\prime L) +
\frac{1}{2g_5^2}\Big[2\Pi^+_g + s_\alpha^2\big(\Pi^-_g -\Pi^+_g\big)  \Big]  \nn \,,
\eea
where $s_\alpha\equiv \sin(\alpha)$ and  $\alpha$ is defined as in eq.(\ref{alphaDef}).
According to eqs.(\ref{defggprime}),  the SM gauge couplings constants and Higgs VEV $v$ are 
\be
\frac{1}{g^2} = \frac{L \Big(5+ c_{2\alpha}+6\theta\Big)}{6g_5^2}\,, \  \  \
\frac{1}{g^{\prime 2}} = \frac{L(1+\theta^\prime)}{g_{5X}^2} + \frac{L (5+c_{2\alpha})}{6g_5^2}, \ \  \
v^2 = \frac{2 s_\alpha^2}{g_5^2 L}=f_\pi^2 s_\alpha^2,
\label{defgso5}
\ee
where $c_{2\alpha}=1-2s_\alpha^2$. 
We immediately see from eq.(\ref{Piso5}) that $\hat T=0$. This is of course not a coincidence, but
a consequence of the custodial $SU(2)_D$ symmetry which is unbroken at $y=L$ \cite{Agashe:2003zs}.
We also have, at tree-level,
\bea
\hat S \a= \a\frac{2s_\alpha^2}{5+c_{2\alpha} +6\theta} \,, \ \ \ \ \ 
W = \frac{(23+7 c_{2\alpha})s_\alpha^2}{5(5+c_{2\alpha}+6\theta)^2}\,, \nn \\ 
Y\a=\a \frac{5(5+c_{2\alpha}+6\theta) +\tan^2 \theta_W \big[-2+23\theta^\prime + c_{2\alpha}(2+7\theta^\prime)\big]}{5(1+\theta^\prime)(5+ c_{2\alpha} +6\theta)^2} s_\alpha^2\,.
\label{STWYso5}
\eea
For $v\ll f_\pi$, we can expand the third relation in (\ref{defgso5}) and find
the correct SM limit $\langle h\rangle  \simeq v$.
When $\theta\sim \theta^\prime \gg1 $, the universal parameters (\ref{STWYso5})
are suppressed. More specifically $\hat S\propto 1/\theta$, $W\sim Y\approx 1/\theta^2$, so that
$\hat S$ becomes the main parameter to keep under control. For large $\theta$, the
mass of the $W$ is given by
\be
M_W \simeq \frac{s_\alpha}{\sqrt{2} L \sqrt{\theta}}\,.
\label{mwSO5}
\ee
Plugging eq.(\ref{mwSO5}) back in eq.(\ref{STWYso5}) gives $\hat S = 2 m_W^2 L^2/3$, like in models with
a bulk Higgs, eq.(\ref{stwyHiggsbulk}), and in the SU(3) model, eq.(\ref{obliquepar}).
The total spectrum of vector KK resonances is given by the zeros of $\Pi_{W_1W_1}$ and of
$\Pi_{W_3W_3}\Pi_{BB}-\Pi_{W_3B}^2$. The lightest non-SM vector mesons arise from the KK tower associated to the $W$ bosons. Before EWSB, their masses $M_n$ are given by the non-vanishing zeros of the following equation: 
\be
\theta M_{n} L+\tan (M_n L) = 0\,.
\label{towergauge}
\ee 
For $\theta\gg 1$ eq.(\ref{towergauge}) gives
\be
M_{KK}^g \equiv |M_1| \simeq \frac{\pi}{2L}\,.
\label{mkkgauge}
\ee
It is useful to pause here and see more closely the relation between this model and its relative in a warped RS compactification \cite{Contino:2006qr}.
Eqs.(\ref{Piso5}), being fixed by symmetry considerations,  are still valid, with the form factors being given by (with no localized gauge kinetic terms)
\bea
\Pi_{g,RS}^{+} \a = \a p\frac{J_0(p z_{UV}) Y_0(p z_{IR}) - Y_0(p z_{UV}) J_0(p z_{IR})}{J_1(p z_{UV}) Y_0(p z_{IR}) - Y_1(p z_{UV}) J_0(p z_{IR})} \,, \nn \\
\Pi_{g,RS}^{-} \a = \a p\frac{J_0(p z_{UV}) Y_1(p z_{IR}) - Y_0(p z_{UV}) J_1(p z_{IR})}{J_1(p z_{UV}) Y_1(p z_{IR}) - Y_1(p z_{UV}) J_1(p z_{IR})} \,, 
\eea
where $J$ and $Y$ are Bessel functions, the 5D metric is
\be
ds^2 = e^{-2 k y} \eta_{\mu\nu} dx^\mu dx^\nu - dy^2 = 
\left(\frac{z_{UV}}{z}\right)^2 ( \eta_{\mu\nu} dx^\mu dx^\nu - dz^2)\,,
\ee
with $0\leq y\leq L$, $z_{UV}\leq z \leq z_{IR}$, $z=e^{ky}/k$. It is a simple exercise to show that
$\Pi_{g,RS}^\pm \rightarrow \Pi_g^\pm$ as $k\rightarrow 0$, as it should.
The definitions (\ref{defggprime}) give now
\be
\frac{1}{g^2} = \frac{z_{UV} \Big(3 c_{2\alpha}-3+16\log \frac{z_{IR}}{z_{UV}}\Big)}{16g_5^2}\,, \  \  \
\frac{1}{g^{\prime 2}} = \frac{1}{g^2}+
\frac{z_{UV} \log \frac{z_{IR}}{z_{UV}}}{g_{5X}^2}, \ \  \
v^2 = \frac{4 z_{UV}s_\alpha^2}{g_5^2 z_{IR}^2},
\label{defgso5warped}
\ee
so that 
\be
M_W^{(RS)} \simeq \frac{s_\alpha}{z_{IR}\sqrt{ \log \frac{z_{IR}}{z_{UV}}}}\,.
\label{mWwarped}
\ee
For $z_{IR}/z_{UV}> 10^5$, the mass of the first KK vector resonance is roughly fixed to be 
\be
M_{KK}^{g(RS)} \simeq \frac{5}{2z_{IR}}.
\label{mkkwarped}
\ee
Matching  eqs.(\ref{mwSO5}) and (\ref{mkkgauge}) with (\ref{mWwarped}) and (\ref{mkkwarped}), respectively,  gives
\be
\theta \simeq \frac{25}{2\pi^2}\log \frac{z_{IR}}{z_{UV}}\,, \ \ \ \frac 1L \simeq \frac 5\pi \frac 1z_{IR}\,,
\label{matchingwarpedflat}
\ee
providing a precise relation between the warped and the flat model parameters. 
Notice that for warped RS models that aim to solve the hierarchy problem, $\log (z_{IR}/z_{UV})\sim 35$ corresponding 
to $\theta\simeq 44$,  on the edge of perturbativity (see eq.(\ref{boundtheta}) below).
In presence of large localized kinetic terms, the coupling of  KK resonances with states localized at $y=0$
are suppressed, since the KK wave-functions are peaked towards the $y=L$ boundary \cite{Carena:2002me}.
This is exactly what happens in warped space, where the KK resonances are peaked at the IR brane, showing
again the analogies between the RS warped model and the flat one with large localized kinetic terms.

Let us now turn to the fermion sector. 
The SM quarks are embedded in bulk fermions transforming in the
fundamental representation of $SO(5)$, $\mathbf{5}=(2,2)\oplus (1,1)$.
For each quark generation, we introduce 4 bulk fermions $\xi_{q_1}$, $\xi_{q_2}$, $\xi_u$ and $\xi_d$
in the $\mathbf{5}$. Their b.c. are as follows:
\begin{eqnarray}
\begin{array}{c}
\begin{matrix}
\xi_{q_1}=
\begin{bmatrix}
(2,2)^{q_1}_L =
\begin{bmatrix}q^\prime_{1L}(-+) \\ q_{1L}(++) \end{bmatrix}
&
(2,2)^{q_1}_R =
\begin{bmatrix}q^\prime_{1R}(+-) \\ q_{1R}(--) \end{bmatrix}
\\
(1,1)^{q_1}_L(-,-)
&
(1,1)^{q_1}_R(+,+)
\end{bmatrix}_{2/3},
\\  \\
\xi_{q_2}=
\begin{bmatrix}
(2,2)^{q_2}_L =
\begin{bmatrix}q_{2L}(++) \\ q^\prime_{2L}(-+) \end{bmatrix}
&
(2,2)^{q_2}_R =
\begin{bmatrix}q_{2R}(--) \\ q^\prime_{2R}(+-) \end{bmatrix}
\\
(1,1)^{q_2}_L(-,-)
&
(1,1)^{q_2}_R(+,+)
\end{bmatrix}_{-1/3},
\\ \\
\begin{array}{cc}
 \xi_u=
\begin{bmatrix}
(2,2)^u_L(+-)
&
(2,2)^u_R(-+)
\\
(1,1)^u_L(-+)
&
(1,1)^u_R(+-)
\end{bmatrix}_{2/3},
&
\xi_d=
\begin{bmatrix}
(2,2)^d_L(+-)
&
(2,2)^d_R(-+)
\\
(1,1)^d_L(-+)
&
(1,1)^d_R(+-)
\end{bmatrix}_{-1/3}.
\end{array}
\end{matrix}
\end{array}
\label{fieldcomponents}
\end{eqnarray}
We have displayed the field content according to their $SO(4)$ decomposition. 
The subscripts $2/3$ and $-1/3$ denote the $U(1)_{X}$ charge of each multiplet. 
Note that the choice of parities allow for two SM doublet zero
modes, coming from $q_{1L}$ and $q_{2L}$.  We can get rid of one linear combination
by coupling it  with a very large mass mixing term $\epsilon$ to a chiral fermion  doublet $\eta_R$ localized at $y=0$ with $Q_Y=1/6$. The $O(4)\times U(1)_X$ symmetry at $y=L$ allows for the following mass mixing
terms:
\begin{equation}
\tilde m_u \overline{(2,2)}^{q_1}_L (2,2)^u_R
+ \tilde M_u \overline{(1,1)}^{q_1}_R (1,1)^u_L
+ \tilde m_d \overline{(2,2)}^{q_2}_L (2,2)^d_R
+ \tilde M_d \overline{(1,1)}^{q_2}_R (1,1)^d_L
+\mathrm{h.c.}
\label{masstermsso5}
\end{equation}
Leptons are similarly introduced, the only difference being the  $U(1)_X$ charges, being now
$0$ for $\xi_{q1}$ and $\xi_u$, and $-1$ for $\xi_{q_2}$ and $\xi_d$.
We choose as holographic fermion field components $\xi_{q_{1}L}$,  $\xi_{q_{2}L}$,  $\xi_{uR}$ and
$\xi_{dR}$. For non-vanishing mass terms at $y=L$ there is no need to introduce Lagrange multiplier
fermion fields to describe the right-handed zero-mode singlet components coming from $\xi_{q_1,q_2}$, since these fields will be created by the holographic fields in  $\xi_{uR,dR}$, as explicitly shown in the Appendix B.
The Higgs fermion couplings are easily computed from the quadratic lagrangian by performing 
the gauge rotation (\ref{SigmaFer}), with $\Sigma$ as in eq.(\ref{Sigmaso5}),
setting to zero the $(-)$ field components at $y=0$.
More explicitly, in the chosen $SO(5)$ basis and $SU(2)_L\times SU(2)_R$ embedding (see Appendix C),
\be
\chi_{q_1L} =\frac{1}{\sqrt{2}}\left(
\begin{matrix}
d_{1L} \\
-i d_{1L} \\
u_{1L} \\
i u_{1L} \\
0 
\end{matrix}\right), \ \
\chi_{q_2L} =\frac{1}{\sqrt{2}}\left(
\begin{matrix}
u_{2L} \\
i u_{2L} \\
-d_{2L} \\
i d_{2L} \\
0 
\end{matrix}\right), \ \
\chi_{uR} =\left(
\begin{matrix}
0 \\
0 \\
0 \\
0 \\
u_R 
\end{matrix}\right), \ \
\chi_{dR} =\left(
\begin{matrix}
0 \\
0 \\
0 \\
0 \\
d_R 
\end{matrix}\right).
\ee
When $\epsilon\gg \sqrt{L}$, the e.o.m. of $\eta_R$ and $q_{1L}-q_{2L}$ give $\eta_R=q_{1L}-q_{2L}=0$, so that we can ignore the former and identify $q_{1L}=q_{2L}\equiv q_L$ in the holographic Lagrangian. After straightforward but lengthy algebra, we get
\bea
{\cal L}_H\a=\a\bar q_L \frac{\pslash}{p}\bigg[ \Pi_0^q  +s_\alpha^2
\Big(\Pi_1^{q_u} \frac{H^c (H^c)^\dagger}{H^\dagger H}+ \Pi_1^{q_d} \frac{H H^\dagger}{H^\dagger H} \Big)\bigg] q_L
+ \sum_{a=u,d} \bar a_R  \frac{\pslash}{p} \Big(\Pi_0^a +s_\alpha^2 \Pi_1^a \Big) a_R\nn \\
\a+\a \frac{s_{2\alpha}}{2h}(\Pi^u_M \bar q_L H^c u_R +\Pi^d_M \bar q_L H d_R+h.c.) \,,
\label{Lholofinalso5}
\eea
where 
\be
H =\frac{1}{\sqrt{2}} \left( \begin{matrix}
h_1-i h_2 \\
-h_3-i h_4 
\end{matrix}\right)\,, \ \ \ 
H^c \equiv i \sigma_2 H^\star =-\frac{1}{\sqrt{2}} \left( \begin{matrix}
h_3-i h_4 \\
h_1+i h_2 
\end{matrix}\right)\\.
\ee 
The expression of the form factors appearing in (\ref{Lholofinalso5}) is reported in eq.(\ref{Formfactorsso5}).
They give rise to an infinite number of higher derivative operators when expanded for low momenta.  In particular, one has
\be
\Pi^{u,d}_M  \propto (1-\tilde m_{u,d} \tilde M_{u,d})\,.
\ee
The Yukawa couplings vanish when $\tilde m_{u,d} = 1/\tilde M_{u,d}$, because the effective b.c. for the bidoublet and singlet fermion components in the $SO(5)$ multiplet become the same (the localized mass terms become $SO(5)$ invariant) and no fermion Higgs couplings is allowed. In fact, all the terms proportional to $s_\alpha^2$ in eqs.(\ref{Lholofinalso5}) vanish as well in this limit, see eq.(\ref{Formfactorsso5}).
Although $\Pi^{u,d}_M$ are proportional to the localized mass terms and increase when the latter increase, the canonically normalized couplings are maximized for $\tilde m \simeq -1/\tilde M={\cal O}(1)$, since the fermion wave-function renomalization $Z_{L,R}$ increase linearly in the localized mass terms, when the latter are large, see eqs.(\ref{masstop}) and (\ref{massbottom}) below. It is not so illuminating to write down the general formula for the physical SM fermion masses after EWSB, which is quite complicated. Rather, in order to decrease the number of free parameters and be able to write relatively simple analytic expressions, we will focus in the following on the top and bottom quarks, the only relevant fermions in the EWSB process, and on a sub-space of the whole parameter space where we take $\lambda_1 = \lambda_u$,  $\lambda_2>0$ and $\lambda_d<0$, where
$\lambda_{1,2} =LM_{1,2}$,   $\lambda_{u,d} =LM_{u,d}$ are the bulk masses in units of $1/L$.
The signs have been chosen so that the SM doublet $q_L$ and singlet $d_R$ are localized around $y=0$, but other sign choices are possible.  We take $\tilde M_u = -1/\tilde m_u$, to maximize the size of the top Yukawa coupling, and $|\tilde m_d| \sim  |\tilde M_d| \sim 1$. It is worth to emphasize that the above choices of parameters are dictated only by the desire of having a simple analytic description of the model and in no way they should be seen as a tuning. By expanding at leading order in $p$ the above form factors,  we easily get, for $|\lambda_u|, |\lambda_2|, |\lambda_d|\gtrsim 1$, $\theta\gg 1$,
\bea
\frac{M_{top}}{M_W} \a\simeq \a\frac{|\tilde m_u|}{\sqrt{1+\tilde m_u^2}} \frac{4\sqrt{\theta} \lambda_u   e^{-\lambda_u}}{\sqrt{1+\tilde m_u^2+\lambda_u/\lambda_2}}\,, \label{masstop} \\
\frac{M_{bottom}}{M_W} \a\simeq \a \frac{|1-\tilde m_d\tilde M_d|}{|\tilde M_d|} \frac{2\sqrt{\theta}\sqrt{|\lambda_d\lambda_2|}   e^{-(\lambda_2-\lambda_d)}}{\sqrt{1+(1+\tilde m_u^2)\lambda_2/\lambda_u}}\,.
\label{massbottom}
\eea
The $\lambda_u$ ($\lambda_2$) dependence in $M_{top}$ ($M_{bottom}$) is due to the localized fermion $\eta_R$ at $y=0$, needed to get rid of the extra unwanted SM doublet.  
Eqs.(\ref{masstop}) and (\ref{massbottom}) show how a large localized gauge kinetic term parameter $\theta$ nicely solves the top mass problem. As expected from our general arguments, the SM Yukawa couplings are exponentially suppressed and eqs.(\ref{masstop}), (\ref{massbottom}) suggest us to focus on the region in which $|\lambda_2-\lambda_d|>\lambda_u$.

The spectrum of fermion resonances beyond the SM fields is quite rich. In order to get all the KK towers
one has to retain all fermion components that vanish at $y=0$ and introduce Lagrange multipliers for them, and solve
for the vanishing components, as sketched in eqs.(\ref{VferYuk})-(\ref{LholoYukExp}). 
We will not discuss the resulting spectrum in detail here. It is given by zeros and poles of the form factors (\ref{Formfactorsso5}) before EWSB and by suitable combinations of them after EWSB. We just mention that
before EWSB we get KK towers of fermions in ${\bf 2}_{7/6}$,  ${\bf 2}_{-5/6}$,  ${\bf 2}_{1/6}$,  ${\bf 1}_{2/3}$, ${\bf 1}_{-1/3}$ of $SU(2)_L\times U(1)_Y$. 
The lightest particles beyond the SM are the first fermion resonances in the ${\bf 2}_{7/6}$ tower. Their masses are given by the zeros of $D_{q_1u}(\tilde m_u)$ and for $\tilde m_u = 1$ are roughly  given by
 \be
 M_{KK}^f  \simeq  \frac{\sqrt{2}}{L} \lambda_u e^{-\lambda_u} \,. 
 \label{mkk}
 \ee 
Eq.(\ref{mkk}) puts an upper bound on the values of $\lambda_u$ one could take, otherwise unwanted ultra-light
fermions appear. 

We are now ready to better quantify the relevant region in parameter space that should be considered.
Having understood that $\theta\gg 1$ is the most promising region, the phenomenological requirement 
\be
\hat S\simeq \frac 23 M_W^2 L^2 \simeq \frac 13 \frac{s_\alpha^2}{\theta} \approx 10^{-3}
\ee
fixes $L^{-1} \gtrsim 1 {\rm TeV}$ and  $s_\alpha/\sqrt{\theta}\leq {\cal O}(10^{-1})$. The key parameter determining how much the effective potential should be tuned to give a small $s_\alpha$ is hence  $\theta$. Larger the latter is, more natural the model is. The drawback is that larger $\theta$ correspond to stronger $5D$ couplings, since, at fixed 4D coupling $g$, eq.(\ref{defgso5}) shows that $g_5^2$ has to increase linearly with $\theta$.
An order of magnitude estimate on the allowed range of $\theta$ is provided by NDA, applied to the 5D coupling constant $g_5$. According to NDA, perturbativity in the effective 5D theory is lost at energies $E$
when\footnote{Notice the appearance in eq.(\ref{NDA}) of the $4D$ phase factor $16\pi^2$ rather than the 5D one $24\pi^3$, as often (too optimistically) taken in the literature. This is due to the fact that in odd dimensions an extra factor of $\pi$ generally arises from the loop integral. It can be explicitly verified by performing a one-loop computation using, say, Pauli-Villars regularization.}
\be
\frac{g_5^2 E}{16\pi^2} \sim 1 \Rightarrow E L \sim \frac{16\pi^2}{g^2 \theta}\,.
\label{NDA}
\ee
Requiring $E L\gg 1$ gives $\theta \ll 400$. A more precise estimate would also take into account the multiplicity of fields, which typically tend to lower the range of validity of the theory, so that a more conservative and realistic bound would approximately be  
\be
\theta \ll 10^2\,.
\label{boundtheta}
\ee

Having roughly fixed the size of some of the crucial parameters in the model, we have now to check whether
EWSB occurs or not in this parameter range. We then turn our attention to the one-loop Higgs effective potential. The gauge contribution to the Higgs potential, for $\theta\sim \theta^\prime \gg 1$ and $s_\alpha \ll 1$, is well approximated by
\be
V_g \simeq \frac{3}{2} \int \!\!\! \frac{d^4p}{(2\pi)^4} \bigg[ 2\log\bigg(1+s_{\alpha}^2\frac{ \Pi_g^--\Pi_g^+}{2(\Pi_g^++\theta L p^2)}\bigg)+\log\bigg(1+ s_{\alpha}^2\frac{\sec^2 \theta_W ( \Pi_g^--\Pi_g^+)}{2 (\Pi_g^++\theta L p^2)}\bigg)\bigg]\,.
\label{Potso5gauge}
\ee
The fermion contribution $V_f=V_u+V_d$ is the sum of the KK towers associated to the up and down contributions, easily derived from eq.(\ref{Lholofinalso5}): 
\bea
V_u \a= \a -2 N_c \int  \!\!\! \frac{d^4p}{(2\pi)^4}  \log\bigg[\Big(1+s_{\alpha}^2\frac{ \Pi_1^{q_u}}{\Pi_0^q}\Big)\Big(1+s_{\alpha}^2\frac{ \Pi_1^{u}}{\Pi_0^u}\Big)-s_{2\alpha}^2\frac{(\Pi^u_M)^2}{8\Pi_0^q\Pi_0^u}\bigg] \,, \label{VfUp}\\
V_d \a= \a   -2 N_c \int  \!\!\! \frac{d^4p}{(2\pi)^4}  \log\bigg[\Big(1+s_{\alpha}^2\frac{ \Pi_1^{q_d}}{\Pi_0^q}\Big)\Big(1+s_{\alpha}^2\frac{ \Pi_1^{d}}{\Pi_0^d}\Big)-s_{2\alpha}^2\frac{(\Pi^d_M)^2}{8\Pi_0^q\Pi_0^d}\bigg]\  \,, \label{VfDown}
\eea
where $N_c=3$ is the QCD color factor.\footnote{Eqs.(\ref{Potso5gauge}) and (\ref{VfUp}) are very similar to eqs.(B.5) and (B.9) of \cite{Panico:2008bx}, where a modified version of the MCHM$_5$ model was considered, expressed in terms of similar form factors.}
The total Higgs potential is given by $V=V_g+V_u+V_d$. For generic values of the input parameters, $\theta$, $\theta^\prime$, $\lambda$'s and $\tilde m$'s, it is quite hard to get a reliable and sufficiently treatable analytic approximation
for $V$. However, when $\tilde M_u = -1/\tilde m_u$, a great simplification occurs,
because $\Pi_1^{q_u}=\Pi_1^u = 0$. Given the lightness of the bottom quark, the form factor $\Pi^d_M$ can safely be neglected and the total potential has the form  (\ref{PotApprox2}), with $c=c_d+c_g$, $d=d_u$ and
\bea
c_g \a = \a  \frac 32 L^4\int \!\!\! \frac{d^4p}{(2\pi)^4}  \frac{\Pi_g^--\Pi_g^+}{ \Pi_g^++\theta L p^2}  \Big(1+\frac{\sec^2\theta_W}{2}\Big)
\,, \nn \\
c_d\a= \a -2N_cL^4 \int  \!\!\! \frac{d^4p}{(2\pi)^4} \Big(\frac{\Pi_1^{q_d}}{\Pi_0^q}+\frac{\Pi_1^{d}}{\Pi_0^d}\Big)\,, \nn \\
d_u\a= \a -\frac{N_c}4 L^4 \int  \!\!\! \frac{d^4p}{(2\pi)^4} \frac{(\Pi^u_M)^2}{\Pi_0^q\Pi_0^u}\,.
\eea
The Higgs mass is approximatively given by 
\be
M_H \simeq \frac{M_W g\theta}{\alpha_0} \sqrt{V^{\prime\prime}(\alpha_0)} \simeq
2M_W \theta g\sqrt{4d+c} \,.
\label{mhiggsapp}
\ee
The loop factor coming from the square root term in eq.(\ref{mhiggsapp}) is more than compensated by the
factor $\theta$, so that the LEP bound on $M_H$ is easily evaded.
It is straightforward to numerically check the existence of wide ranges in the input parameters where $|c/d|\leq 4$,
such that $s_\alpha$ is small enough and the main phenomenological bounds, such as 
$\hat S\sim 10^{-3}$, $M_H> 114$ GeV,  correct top and bottom masses, are fulfilled.\footnote{Notice the importance of having approximate analytic formulae for $M_{top}$ and $M_{bottom}$ that do not depend on $s_\alpha$. The latter, indeed, is fixed by the total potential $V$ and depends on all the input parameters of the model in a complicated way. On the other hand, without knowing $s_\alpha$ and without formulae like eqs.(\ref{masstop}) and (\ref{massbottom}), there would be no way to fix (some of) the input parameters in the model and the only practical way to proceed would be by means of numerical random scans in the parameter space.}
In order to restrict the parameter space region to study, we may proceed as follows. 
We maximize the top Yukawa coupling by taking $\tilde M_u = -1/\tilde m_u = 1$ and fix the localized gauge kinetic terms to  $\theta=\theta^\prime = 25$. The top mass relation (\ref{masstop}) fixes then $\lambda_u$ to lie in a narrow range
$\lambda_u\simeq 2$, depending only mildly on $\lambda_2$. Fixing $\lambda_u$ to an arbitrary value close to 2
will fix $\lambda_2$. Given $\tilde m_d$ and $\tilde M_d$, the bottom mass formula (\ref{massbottom}) will fix $\lambda_d$, so that
we are left with a 3 parameter space spanned by $(\tilde m_d, \tilde M_d)$ and $\lambda_u$ around the value 2.
As an example, let us work out a specific set of parameters given by  $\tilde m_d = -2/5$, $\tilde M_d = 1/5$, $\lambda_u = 2.18$. Eqs.(\ref{masstop}) and (\ref{massbottom}) fix $\lambda_2\simeq 3.16$,  $\lambda_d \simeq - 4.47$.
For such input parameters, we get
\bea
\a\a s_\alpha \simeq \frac 13 \,, \ \ \ 
\hat S \simeq 1.4\times 10^{-3}, \ \ \ \frac 1L\simeq 1.8 \;{\rm TeV}\,,\nn \\
\a\a   M_H \simeq 130 \;{\rm GeV} \,,\ \  \ M_{KK}^f \simeq 630\; {\rm GeV}\,,\ \ \  M_{KK}^g \simeq 2.6 \;{\rm TeV} \,.
\eea
The fine-tuning associated to this specific model can easily be computed using eq.(\ref{finetuning}). The dominant sensitivity is in the $\lambda_d$ and $\lambda_2$ directions. By numerically computing eq.(\ref{finetuning}), we get a modest tuning around $10\%$.

By appropriately choosing the bulk mass parameters, as well as by introducing large localized fermion kinetic terms as well, 
we can localize all the remaining light SM fermion fields sufficiently close to $y=0$, so that universality of the gauge couplings is achieved with the required accuracy. We will not try to make a quantitative matching between the flat space model and the corresponding Randall-Sundrum (RS) warped one \cite{Contino:2006qr} in the fermion sector, as briefly done in the gauge sector. 
Similarly,  we will not address here other important bounds that should be taken into account, such as $\delta g_b$ or
a more carefully analysis of the universal oblique corrections up to one-loop level, particularly important for the $\hat T$ parameter. We leave a more detailed analysis of these promising $SO(5)$ flat models with large gauge kinetic terms for future investigations.

\subsection{Higher dimensional models}

One compact extra dimension is the minimal scenario where most
progress has been achieved so far. Even if no (semi-)realistic non-supersymmetric GHU model
in more than one extra dimension has been found, it is worth to briefly see what  are the new qualitative features that one encounters in
more extra dimensions. Since the NDA estimate of the cut-off $\Lambda$ in higher dimensional theories 
decreases as the number of extra dimensions increase and no new fundamental features seem to appear
in further increasing their number, let us only consider the case
of two extra dimensions, namely theories in 6 space-time dimensions.
In 6D, there are several potentially interesting two-dimensional compact spaces
one could consider. The simplest spaces, leading to a 4D chiral spectrum of fermions,
are given by orbifolds of tori of the form $T^2/\Z_N$, where $N=2,3,4,6$.
Let us focus on these spaces in the following.

There are two main important qualitative features that happen when going to 6D.
The first, good feature, is the appearing of a gauge-invariant Higgs quartic coupling at tree-level,
arising from the non-abelian part of the internal components of the
gauge field kinetic term.
A tree-level quartic coupling is welcome, because it can automatically
solve the problem of a too light Higgs without the need of introducing extra complications.
The second, bad feature, is the possible appearance of a local, gauge-invariant, operator
that contributes to the Higgs mass. This is an operator localized at the fixed-points of the
$T^2/\Z_N$ orbifold, with a quadratically divergent coefficient,
in general \cite{CGM,vonGersdorff2,SSSW} (see also \cite{Biggio:2004kr} for an analysis in $D>6$ dimensions).
It is linear in the internal components of the field-strength $F$. Its abelian term
corresponds to a tadpole for certain gauge field components, whereas its non-abelian part
represents a mass term for the Higgs field. If there is no symmetry to get rid
of this operator, the hierarchy problem is reintroduced.
It turns out that  in 6D a discrete symmetry forbidding this operator can be implemented
only for $T^2/\Z_2$ orbifolds, in which case, however, one gets two Higgs doublets, rather than one.
In this situation, the Higgs effective potential has various similarities with the one
arising in the Minimal Supersymmetric Standard Model. Explicit computations on
a given 6D model \cite{Antoniadis:2001cv} have shown that the lightest Higgs field turns
out to be again too light \cite{Hosotani:2004wv}.

Maybe a more interesting possibility is obtained by considering $T^2/\Z_N$ orbifolds, with $N\neq 2$.
If $N\neq 2$, one can get 2, 1 or 0 Higgs doublets, depending on the orbifold projection.
The most interesting case appears to be given by the 1 Higgs doublet models,
for which one finds $M_H = 2 M_W$ at tree-level, by geometrical considerations \cite{SSSW}.
However, no symmetry forbids the appearance of the localized operator mentioned above, which would
spoil the stabilization of the electroweak scale. Even if this operator is put to zero at tree-level,
no accidental one-loop cancellation seems to be possible. The best one can do is to advocate
a spectrum of 6D fields such that the sum of the one-loop quadratically divergent coefficients over all
fixed points vanish (global cancellation). In this case, it actually turns out that the
electroweak scale is not destabilized. Contrary to the 5D constructions considered before,
the quadratic sensitivity to the cut-off would presumably be reintroduced at two-loop level, 
but a one-loop cancellation is enough to solve the little hierarchy problem. 

\section{Conclusions}

Quantum field theories in extra dimensions are a promising arena for new physics beyond the SM, 
in particular to address the so far mysterious EWSB mechanism in the SM. 
Natural models arise from 5D theories defined on a segment
where the Higgs field is identified with the internal components of a gauge field. 
I have reviewed here the basics of the holographic method to technically deal with such (and other) theories,
and then applied it to the construction of two simple models based on $SU(3)$ and $SO(5)$ electroweak gauge groups, respectively.
The $SO(5)$ model is generally more promising and natural than the $SU(3)$ one, but the latter is  more weakly coupled. 
In fact, the  $SO(5)$ model of section 4.2, like their warped GHU analogues, is at the edge of calculability.

Although GHU models in warped spaces have the important additional features of explaining how the TeV scale dynamically
arises  (issue which is not addressed in flat space, where TeV$^{-1}-$sized extra dimensions are taken for granted) and
can also allow for a calculable theory of flavour, the LHC TeV-physics associated to the EWSB mechanism is essentially the same in warped or flat space. Roughly speaking, the lightest non-SM particles predicted are always colored spin 1/2 resonances with similar quantum numbers and interactions. Their mass can be well below
the TeV scale, as it happens, for instance, in the SO(5) model, and hence visible at the LHC. 
Models in flat space, as we have briefly sketched in subsection 4.2, can also be seen as effective simple descriptions of warped models,
when large localized kinetic terms are inserted, and allow more flexibility.

\section*{Acknowledgments}

I would like to thank Giuliano Panico for discussions and INFN for partial financial support.

\appendix

\section{Conventions}
\label{conventions}

We work in the ``mostly minus" convention for the 5D metric:
\be
ds^2 = \eta_{\mu\nu}dx^\mu dx^\nu -dy^2 = (dx^0)^2 - (d\vec x)^2 - dy^2.
\ee
We always denote by $x^\mu$  the four space-time dimensions, with $\mu=0,1,2,3$.
The internal coordinate is parametrized by $y$, ranging from $0$ to $L$, with $L$ the length of the segment.
Five-dimensional indices are denoted by capital latin letters $M,N,\ldots$, with $M=(\mu,y)$.
Correspondingly, 5D vectors decompose as $A_M=(A_\mu,A_y)$ under the $SO(4,1)\rightarrow SO(3,1)$
decomposition. The 5D gamma matrices are taken as $\gamma^M=(\gamma^\mu,\gamma^y) = (\gamma^\mu,-i \gamma^5)$, 
with $(\gamma^5)^2=1$. Left-handed and right-handed fermions $\psi_L$ and $\psi_R$ are defined as
$\gamma^5 \psi_{L} = - \psi_{L}$, $\gamma^5 \psi_{R} = + \psi_{R}$.

Neumann and Dirichlet boundary conditions for the fields are schematically denoted as $(+)$ and $(-)$.
For brevity, we report together the b.c. at $y=0$ and at $y=L$ of any field by writing $(\pm \pm)$, 
 with the first and second entries referring to $y=0$ and $y=L$, respectively.
The Fourier transform of the fields are always denoted with the same letter as the field themselves.
Finally, in order to avoid confusion between the different mass terms that can appear in 5D theories, we
denote by lower letters $m$ the 5D bulk mass terms, by capital letters $M$ the mass eigenvalues of 4D fields
and by a tilde $\tilde m$ or $\tilde M$ 5D mass terms localized at the boundaries of the interval.

\section{Mass terms at $y=L$}

An interesting class of models are obtained by introducing localized fermion mass terms at $y=L$, so 
we analyze in detail this case. 
Consider a pair of bulk fermions $\psi_1$ and $\psi_2$, mixed through localized mass terms as follows: 
\be
{\cal L}= \int_0^L\!dy \sum_{j=1,2} \bigg[\frac i2 \bar \psi_j \gamma^M \partial_M \psi_j - \frac i2 (\partial_M \bar \psi_j) \gamma^M \psi_j - m_j \bar \psi_j \psi_j \bigg] + \tilde m(\bar \psi_{1L}\psi_{2R}+\bar\psi_{2R}\psi_{1L})(L)\,,
\ee
where $\tilde m$ is a dimensionless mass parameter (being $\psi$  a 5D field, $[\bar \psi \psi]=4$).  Let us take $\chi_{L}=\psi_{1L}(0)$ and $\chi_R=\psi_{2R}(0)$ as holographic fields. As discussed in the main text, the vanishing of the boundary variations at $y=0$ require the addition of the following term at $y=0$:
\be
{\cal L}_{0}= \frac 12 \Big( \bar \psi_{1L} \psi_{1R} + \bar \psi_{1R} \psi_{1L}\bigg)-
 \frac 12 \Big( \bar \psi_{2L} \psi_{2R} + \bar \psi_{2R} \psi_{2L}\bigg)\,.
\label{L0fer2}
\ee
Due to the localized mass terms, the boundary variations at $y=L$ is not automatically vanishing now, and the following term at $y=L$ has to be added:
\be
{\cal L}_{L}= -\frac 12 \Big( \bar \psi_{1L} \psi_{1R} + \bar \psi_{1R} \psi_{1L}\bigg)+
 \frac 12 \Big( \bar \psi_{2L} \psi_{2R} + \bar \psi_{2R} \psi_{2L}\bigg)
\label{L0fer2L} \,,
\ee
which give 
\bea
\psi_{1R}(L) \a = \a \tilde m\, \psi_{2R}(L) \,, \nn \\
\psi_{2L}(L) \a = \a - \tilde m\, \psi_{1L}(L)\,,
\label{bcmterms}
\eea
as effective b.c. at $y=L$. After a simple computation, we  get
\bea
\psi_{1L} (y)\a = \a \frac{\Big[G_+(-m_2) G_+(y,m_1) -\tilde m^2 G_-(m_2) G_-(y,m_1)\Big] \chi_L +\tilde m \omega_2 G_-(L-y,m_1)\frac{\pslash}{p} \chi_R}
{ G_+(m_1)G_+(-m_2)-\tilde m^2 G_-(m_1) G_-(m_2) }, \nn \\
\psi_{1R} (y)\a = \a \frac{\Big[G_+(-m_2) G_-(y,m_1) +\tilde m^2 G_-(m_2) G_+(y,-m_1)\Big]\frac{\pslash}{p} \chi_L +\tilde m \omega_2 G_+(L-y,m_1) \chi_R}{ G_+(m_1)G_+(-m_2)-\tilde m^2 G_-(m_1) G_-(m_2) }, \nn \\
\psi_{2L} (y)\a = \a -\frac{\Big[G_+(m_1) G_-(y,m_2) +\tilde m^2 G_-(m_1) G_+(y,m_2)\Big]\frac{\pslash}{p} \chi_R +\tilde m \omega_1 G_+(L-y,-m_2) \chi_L}{ G_+(m_1)G_+(-m_2)-\tilde m^2 G_-(m_1) G_-(m_2) }, \nn \\
\psi_{2R} (y)\a = \a \frac{\Big[G_+(m_1) G_+(y,-m_2) -\tilde m^2 G_-(m_1) G_-(y,m_2)\Big] \chi_R +\tilde m \omega_1 G_-(L-y,m_2)\frac{\pslash}{p} \chi_L}
{ G_+(m_1)G_+(-m_2)-\tilde m^2 G_-(m_1) G_-(m_2) }. \nn
\label{ferpropmassL}
\eea
For $\tilde m =0$, they reduce to
\be
\left\{
\begin{array}{ll}
\ds\psi_{1L}(y)= \frac{G_+(y,m_1)}{G_+(m_1)} \chi_L\,, \a  \ds\psi_{2L}(y)=- \frac{G_-(y,m_2)}{G_+(-m_2)} \frac{\pslash}p \chi_R\,,\\
\ds\psi_{1R}(y)=\frac{G_-(y,m_1)}{G_+(m_1)}\frac{\pslash}p \chi_L\,, \a  \ds\psi_{2R}(y)=\frac{G_+(y,-m_2)}{G_+(-m_2)} \chi_R\,,
\end{array}
\right. 
\label{GfermionchiLN}
\ee
while for $\tilde m\rightarrow \infty$
\be
\left\{
\begin{array}{ll}
\ds\psi_{1L}(y)= \frac{G_-(y,m_1)}{G_-(m_1)} \chi_L\,, \a  \ds\psi_{2L}(y)= \frac{G_+(y,m_2)}{G_-(m_2)} \frac{\pslash}p \chi_R\,,\\
\ds\psi_{1R}(y)=-\frac{G_+(y,-m_1)}{G_-(m_1)}\frac{\pslash}p \chi_L\,, \a  \ds\psi_{2R}(y)=\frac{G_-(y,m_2)}{G_-(m_2)} \chi_R\,.
\end{array}
\right. 
\label{GfermionchiLNinf}
\ee
As can be seen, due to $\tilde m$, $\psi_1$ ($\psi_2$) has a non-vanishing overlap with the holographic component $\chi_R$ ($\chi_L$) of $\psi_2$ ($\psi_1$). This is small
for $\tilde m\ll 1$ and $\tilde m\gg 1$ and it is maximal for $\tilde m= {\cal O}(1)$.
As happens in the scalar case, for very large $\tilde m$, we effectively flip the b.c. of all fermion components. 

The holographic Lagrangian can be written as
\be
{\cal L}_H = \bar \chi_L \;\frac{\pslash}{p} \,\frac{N^-_{21}(\tilde m)}{D_{12}(\tilde m)}  \chi_L +  \bar \chi_R\; \frac{\pslash}{p} \,\frac{N^+_{12}(\tilde m)}{D_{12}(\tilde m)} \chi_R +\frac{M_{12}(\tilde m)}{D_{12}(\tilde m)} \Big(\bar \chi_L \chi_R+\bar \chi_R \chi_L \Big) \,,
\ee
where
\bea
N_{ij}^\pm(\tilde m) \a = \a G_+(\pm m_i)  G_-(m_j)+ \tilde m^2 G_-(m_i)  G_+(\pm m_j)\,, \nn \\
D_{12}(\tilde m)\a = \a G_+(m_1) G_+(-m_2)-\tilde m^2 G_-(m_1) G_-(m_2) \,, \nn \\
M_{12} (\tilde m) \a = \a  \tilde m \omega_1 \omega_2 \,.
\label{FunctionsNpmD}
\eea
Similarly, we could have considered the other choice of localized mass term, namely
\be
\tilde m(\bar \psi_{1L}\psi_{2R}+\bar\psi_{2R}\psi_{1L})(L)\rightarrow
\tilde m(\bar \psi_{1R}\psi_{2L}+\bar\psi_{2L}\psi_{1R})(L)\,.
\ee
In that case, the localized term to be added at $y=L$ is the opposite of eq.(\ref{L0fer2L}), and the resulting b.c.  are 
\bea
\psi_{2R}(L) \a = \a \tilde m\, \psi_{1R}(L) \,, \nn \\
\psi_{1L}(L) \a = \a - \tilde m\, \psi_{2L}(L)\,,
\eea
namely as in eq.(\ref{bcmterms}), but with $\tilde m \rightarrow 1/\tilde m$. Keeping the same holographic fields as before, $\chi_{L}=\psi_{1L}(0)$ and $\chi_R=\psi_{2R}(0)$, one has
\be
{\cal L}_H = \bar \chi_L \;\frac{\pslash}{p} \,\frac{N^-_{21}(1/\tilde m)}{D_{12}(1/\tilde m)}  \chi_L +  \bar \chi_R\; \frac{\pslash}{p} \,\frac{N^+_{12}(1/\tilde m)}{D_{12}(1/\tilde m)} \chi_R +\frac{M_{12}(1/\tilde m)}{D_{12}(1/\tilde m)} \Big(\bar \chi_L \chi_R+\bar \chi_R \chi_L \Big) \,.
\ee

\section{$SO(5)$ generators and fermion form factors}

We list here the explicit choice of $SO(5)$ generators and $SU(2)_L\times SU(2)_R$ embedding used in section 4.
Denoting by 
\be
t^{ab}_{ij} = - t^{ba}_{ij} =  \delta_i^a \delta_j^b -  \delta_i^b \delta_j^a
\ee
the 10 anti-symmetric generators of $SO(5)$, where $a,b=1,\ldots,5$ label the generators
and $i,j$ their matrix components, we take
\bea
t^1_L \a= \a -\frac i2 (t^{23}+t^{14}), \ \ t^2_L = -\frac i2 (t^{31}+t^{24}), \ \ t^3_L = -\frac i2 (t^{12}+t^{34}), \ \  \nn \\
t^1_R\a= \a -\frac i2 (t^{23}-t^{14}), \ \ t^2_R = -\frac i2 (t^{31}-t^{24}), \ \ t^3_R = -\frac i2 (t^{12}-t^{34}), \ \  \nn \\
t^{\hat a} \a = \a  -\frac{i}{\sqrt{2}} t^{a5}, \ \ \hat a = 1,2,3,4 \,.
\eea
In this basis, $t^{1,2,3}_L$ generate $SU(2)_L$, $t^{1,2,3}_R$ generate $SU(2)_R$ and
$t^{\hat 1,\hat 2,\hat 3,\hat 4}\in SO(5)/SO(4)$. 

In terms of the functions (\ref{FunctionsNpmD}),  the form factors appearing in  the holographic fermion Lagrangian
(\ref{Lholofinalso5}) are the following:
\bea
\Pi_0^q \a= \a  \frac{N_{uq_1}^-(\tilde m_u)}{D_{q_1u}(\tilde m_u)}  +  \frac{N_{dq_2}^-(\tilde m_d)}{D_{q_2d}(\tilde m_d)}  \,, \nn \\
\Pi_1^{q_u} \a = \a  \frac 12 \bigg(  \frac{N_{uq_1}^-(1/\tilde M_u)}{D_{q_1u}(1/\tilde M_u)}  - \frac{N_{uq_1}^-(\tilde m_u)}{D_{q_1u}(\tilde m_u)} \bigg) \,, \ \ \ \
\Pi_1^{q_d}  =  \frac 12 \bigg(  \frac{N_{dq_2}^-(1/\tilde M_d)}{D_{q_2d}(1/\tilde M_d)}  - \frac{N_{dq_2}^-(\tilde m_d)}{D_{q_2d}(\tilde m_d)} \bigg)  \,, \nn \\
\Pi_0^u \a= \a  \frac{N_{q_1u}^+(1/\tilde M_u)}{D_{q_1u}(1/\tilde M_u)} \,, \hspace{2.8cm}
\Pi_1^u  =    \frac{N_{q_1u}^+(\tilde m_u)}{D_{q_1u}(\tilde m_u)} - \frac{N_{q_1u}^+(1/\tilde M_u)}{D_{q_1u}(1/\tilde M_u)} \,, \nn \\
\Pi_0^d \a= \a  \frac{N_{q_2d}^+(1/\tilde M_d)}{D_{q_2d}(1/\tilde M_d)} \,, \hspace{2.8cm}
\Pi_1^d  =    \frac{N_{q_2d}^+(\tilde m_d)}{D_{q_2d}(\tilde m_u)} - \frac{N_{q_2d}^+(1/\tilde M_d)}{D_{q_2d}(1/\tilde M_d)}  \,, \nn \\
\Pi^{u}_M  \a = \a \frac{M_{q_1u}(\tilde m_u)}{D_{q_1u}(\tilde m_u)} - \frac{M_{q_1u}(1/\tilde M_u)}{D_{q_1u}(1/\tilde M_u)} \,, \ \ \ 
\Pi^{d}_M   =   -\frac{M_{q_2d}(\tilde m_d)}{D_{q_2d}(\tilde m_d)} + \frac{M_{q_2d}(1/\tilde M_d)}{D_{q_2d}(1/\tilde M_d)}  \,.
\label{Formfactorsso5}
\eea

\end{document}